\documentclass[12pt]{article}
\usepackage{geometry}
\usepackage{amssymb}
\usepackage{amsmath}
\usepackage{bbm}
\usepackage{MnSymbol}
\usepackage{hyperref}
\usepackage{braket}
\usepackage{cite}
\usepackage{subcaption,tikz}
\usetikzlibrary{arrows,decorations.markings,decorations.pathreplacing}

\newcommand{\be}{\begin{equation}}
\newcommand{\ee}{\end{equation}}
\newcommand{\bea}{\begin{eqnarray}}
\newcommand{\eea}{\end{eqnarray}}

\newcommand{\io}{\iota}

\newcommand{\C}{\mathbb{C}}
\renewcommand{\H}{\mathcal{H}}

\newcommand{\lrr}{\longrightarrow}

\newcommand{\Z}{\mathbb{Z}}

\newcommand{\id}{\mathbbm{1}}

\tikzset{
	partial ellipse/.style args={#1:#2:#3}{
		insert path={+ (#1:#3) arc (#1:#2:#3)}
	}
}

\numberwithin{equation}{section}

\begin{document}
	
\begin{flushright}
	MI-HET-789
\end{flushright}
	
\begin{center}

{\large\bf Lower-Form Symmetries}

	\vspace*{0.2in}
	
	Thomas Vandermeulen
	
	\vspace*{0.2in}
	
		{\begin{tabular}{l}
				George P.~and Cynthia W.~Mitchell Institute\\
				for Fundamental Physics and Astronomy\\
				Texas A\&M University\\
				College Station, TX 77843 \end{tabular}}
			
	\vspace*{0.2in}
	
	{\tt tvand@tamu.edu}
	
\end{center}
\pagenumbering{gobble}

When gauging a $(d-1)$-form symmetry in $d$ spacetime dimensions, one formally expects the gauged theory to carry a dual $-1$-form symmetry.  This work focuses on the study of such symmetries, in particular via the spacetime-filling topological operators that implement them, in two-dimensional  field theory (with an eye towards general quantum field theory).  As theories with $(d-1)$-form symmetries are known to be equivalent to direct sums of local theories, we review how gauging a $(d-1)$-form symmetry projects onto a single component in this sum, and explain how gauging the resulting $-1$-form symmetry restores the direct sum.

\newpage
\tableofcontents

\newpage

\section{Introduction}
\label{sec:intro}
\pagenumbering{arabic}

When gauging discrete symmetries in two-dimensional field theory, experience has taught us that we should expect the gauged theory to possess a dual symmetry, which can in turn be gauged to reproduce the original theory \cite{Vafa}.  These dual symmetries are commonly referred to as quantum symmetries, so we will use the two terms interchangeably.  The increasing understanding of the importance of global higher-form symmetries has allowed the notion of quantum symmetry to be generalized -- in $d$ spacetime dimensions, a theory gauged by a $p$-form symmetry is expected to have a quantum $d-(p+2)$-form symmetry \cite{GaiottoKapustinSeibergWillett}.  We can immediately verify that this gels with the original notion of quantum symmetry: when $d=2$, 0-form symmetries are dual to other 0-form symmetries.

Of course, it is well-known that $2d$ theories can exhibit a single type of higher-form symmetry, namely 1-form symmetries.  What is the quantum dual to a two-dimensional 1-form symmetry?  The expression above suggests that the answer is a $-1$-form symmetry.  In fact, in any dimension, the `highest'-form symmetry we can have is a $(d-1)$-form symmetry, and in any dimension its dual comes out as a $-1$-form symmetry.  Is this something we can make sense of?

The above characterization of quantum symmetry stems from the fact that the topological operators which implement a $p$-form symmetry will have (in general non-topological) operators of one dimension lower on which they can end.  In two dimensions, for example, these are local operators known as twist fields.  The quantum symmetry is one that naturally acts on these twist operators.  A $p$-form symmetry is controlled by topological operators of dimension $d-(p+1)$, meaning that its twist operators are of dimension $d-(p+2)$, and their symmetries would be $d-(p+2)$-form symmetries.  A $(d-1)$-form symmetry, however, is controlled by local (pointlike) operators, but now we appear to be in trouble since there is not any obvious notion of a twist operator (which would formally need to be of dimension $-1$) on which the putative $-1$-form symmetry would naturally act.

In the face of such confusion, let us put aside the question of which objects would be charged under such a symmetry and focus entirely on the operators which would implement it.  As already noted, a $p$-form symmetry can be associated to a topological operator of codimension $p+1$.  Again looking to two dimensions, ordinary symmetries can be associated with operators of codimension 1, i.e.~line operators, producing the picture of symmetries in $2d$ as topological defect lines (TDLs).  In two dimensions, there are in total three types of topological operators we could have.  Aside from TDLs, we can sensibly describe dimension 0 local operators and dimension 2 surface operators.  Formally, then, these would be associated with 1-form and $-1$-form symmetries, respectively.  They would also be dual under gauging.  Figure~\ref{fig:2dtable} presents an organized summary of these three types.

\begin{figure}[h]
\begin{tabular}{l l l l}
Symmetry & Associated Topological Operator & Codimension & Dimension \\\hline
$-1$-form & Surface/Background Operator & 0 & 2\\
0-form & Topological Defct Line (TDL) & 1 & 1\\
1-form & Topological Point Operator (TPO) & 2 & 0
\end{tabular}
\caption{Types of symmetry in $2d$ theories.}
\label{fig:2dtable}
\end{figure}

A similar story holds in arbitrary dimension.  There can exist topological point operators (TPOs) which implement $(d-1)$-form symmetries, and one expects these to be dual to spacetime-filling operators (operators of dimension $d$) to which we would associate $-1$-form symmetries.  While the existence of $-1$-form symmetries seems to be a natural consequence of the notion of a quantum dual symmetry, they have received at best minor attention (notable exceptions being \cite[section 2.6]{Yu}, \cite[section 1.5]{CordovaFreedLamSeiberg} and \cite[section 3]{KapustinSeiberg}) compared to the so-called higher-form symmetries, which is usually taken to mean $p$-form symmetries for $p>0$.  This paper will attempt to treat these `lower-form' symmetries on the same footing as their more famous cousins, and in particular we will see how their existence helps fill out the interconnected web of symmetries that exists in any dimension.

\subsection{Basic Notions}

This section provides the concepts which will be taken as a starting point in this work.  As they are each quite important, we collect their basic definitions here for easy reference.

\begin{itemize}
\item \textit{Symmetries as Topological Operators} -- We will regard the existence of topological operators of dimension $d-(p+1)$ as equivalent to the existence of a $p$-form symmetry \cite{FuchsRunkelSchweigert,GaiottoKapustinSeibergWillett}.  If those operators form a group under fusion, the symmetry is said to be group-like.  In any case, they should form a unitary fusion category (of which a group is a special case).

\item \textit{Decomposition} -- A theory with a $(d-1)$-form symmetry, by the above assertion, has a non-trivial spectrum of dimension 0 (local) topological operators \cite{AndoHellermanHenriquesPantevSharpe}.  These operators can be used to split the theory into a direct sum.  This phenomenon is known as decomposition, and the constituent theories in the sum are called universes in the decomposition.

\item \textit{Trivially-Acting Symmetries} -- One way to obtain a theory with a $(d-1)$-form symmetry is to gauge a $(d-2)$-form symmetry that acts trivially \cite{AndoHellermanHenriquesPantevSharpe,PantevRobbinsSharpeVandermeulen}.  That is, the natural action of this symmetry on operators of dimension $d-2$ should be trivial.  One way to see why this should be the case is to notice that a trivially-acting symmetry will include twist fields which are weight zero and therefore topological.  The global symmetry of such a theory can then be regarded as being controlled by a non-trivial mix of TDLs and TPOs -- in the gauged theory these twist fields become genuine local operators (in the sense of \cite{KapustinSeiberg}), giving the theory a standalone $(d-1)$-form symmetry \cite{TopOps}.

\item \textit{Quantum Symmetry} -- When gauging a $p$-form symmetry in $d$ spacetime dimensions, the resulting theory is expected to carry a $d-(p+2)$-form symmetry \cite{Vafa,GaiottoKapustinSeibergWillett}.  This new symmetry of the gauged theory is referred to as the quantum symmetry or dual symmetry with respect to the original.  In the case that the original symmetry was described by a group $G$, the operators which describe the quantum symmetry are labeled by irreducible representations of $G$ \cite{BhardwajTachikawa}.  In the further case that $G$ is abelian, the quantum symmetry is group-like and is given by $\hat{G}$, the Pontryagin dual to $G$.  Gauging a quantum symmetry `undoes' the original gauging, in that the resulting theory is isomorphic to the original theory before any gauging \cite{BrunnerCarquevillePlencner}.

\item \textit{Orbifold Composibility} -- We expect that subsequent gaugings of a theory can be composed \cite{FrohlichFuchsRunkelSchweigert}.  Concretely, assume that we begin with a theory $T$ and gauge some symmetry $G$ to arrive at $T/G$.  Now we choose a symmetry $H$ of $T/G$ and gauge it to arrive at a third theory, $(T/G)/H$.  We will take orbifold composibility to be the assertion that there necessarily exists a symmetry $G'$ of $T$ such that gauging $G'$ has the same effect as gauging $G$ then gauging $H$.  That is, as theories,
\be
T/G'=(T/G)/H.
\ee
Note that $G'$ does not necessarily act effectively on $T$, a point to which we will return in section~\ref{sec:qscomp}.

A way to see that this assertion should be true is to regard our $d$-dimensional QFT as a boundary of a QFT in $d+1$ dimensions \cite{Wen,GaiottoKulp}.  Gaugings of the $d$-dimensional boundary theory can be encoded as boundary conditions on the bulk QFT, and then composibility of such gaugings follows from composibility of their associated boundary conditions.
\end{itemize}

\section{Warmup: Quantum Mechanics}
\label{sec:qm}

We begin with the simplest possible example: $(0+1)d$ quantum field theory, better known as quantum mechanics.  Even in such a restricted setting, there should exist two types of topological operators -- local operators and spacetime-filling line operators, corresponding respectively to 0-form and $-1$-form symmetries.  In some sense this makes quantum mechanics the ideal laboratory, as these are the $(d-1)$-form and $-1$-form symmetries which are meant to exist in any dimension and be dual to each other, so any story we're going to tell involving these two ingredients really ought to hold up here, if anywhere.

The system of choice will be a particle on a ring of radius $R$, with coordinate $q(t)$.  In addition to the usual kinetic term, it will be convenient to deform the action by introducing a theta term as
\be
\label{circact}
S_\theta[q]=\frac{1}{2R^2}\int \dot{q}^2dt+\frac{\theta}{2\pi}\int\dot{q}dt.
\ee
Such a theory has been studied in a similar toy model capacity (though with different goals in mind) in \cite{CordovaFreedLamSeiberg,GaiottoKapustinKomargodskiSeiberg,KikuchiTanizaki}.  This elementary system has a discrete set of wavefunctions given by
\be
\label{circwf}
\ket{n}=\frac{1}{\sqrt{2\pi}}e^{inq(t)}.
\ee
While the theta term does not affect the wavefunctions, it does appear in the energy of each state as
\be
\label{circen}
E_n(\theta)=\frac{1}{2R^2}\left(n-\frac{\theta}{2\pi}\right)^2.
\ee
This system carries the symmetries of its target space (in the sense of a sigma model), the circle $S^1$, which manifests as the Hilbert space forming a representation of $O(2)$.  Let us focus on the $\Z_2$ shift symmetry $q\to q+\pi$.  This is clearly a symmetry of the action (\ref{circact}).  If we take $g$ to implement this shift, the wavefunctions (\ref{circwf}) obey
\be
g\cdot\ket{n}=(-1)^n\ket{n},
\ee
splitting the Hilbert space into invariant (even $n$) and anti-invariant (odd $n$).

\subsection{Gauging}
\label{sec:qmgauging}

\begin{figure}[h]
	\begin{subfigure}{0.5\textwidth}
	\centering
	\begin{tikzpicture}
	\filldraw[black] (1,0) circle (2pt);
	\node at (1.5,0) {$\sigma_1$};
	\draw[thick] (0,0) [partial ellipse=0:360:1cm and 1cm];
	\end{tikzpicture}
	\caption{}
	\label{circws1}
	\end{subfigure}
	\begin{subfigure}{0.5\textwidth}
	\centering
	\begin{tikzpicture}
	\filldraw[black] (1,0) circle (2pt);
	\node at (1.5,0) {$\sigma_g$};
	\draw[thick] (0,0) [partial ellipse=0:360:1cm and 1cm];
	\end{tikzpicture}
	\caption{}
	\label{circws2}
	\end{subfigure}
\caption{Two ways of inserting topological point operators corresponding to a $\Z_2$ symmetry on a 1$d$ worldsheet.}
\label{fig:circws}
\end{figure}
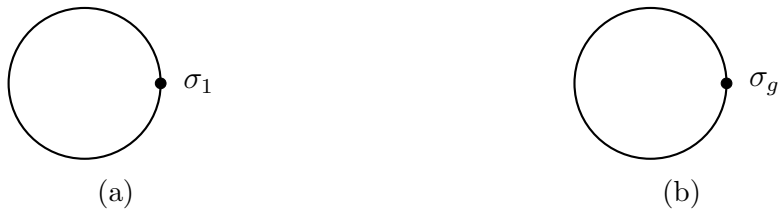

What would the result of gauging this $\Z_2$ symmetry be?  We don't often speak of gauging discrete symmetries in quantum mechanics, but in order to realize the picture laid out in the introduction we would require such a notion.  From the perspective of topological operators, we should be able to gauge by introducing onto the worldsheet a network of local operators associated to the symmetry and then summing over configurations.  In quantum mechanics our worldsheet is the one-dimensional worldline, which for convenience we will compactify to a circle, done by imposing periodic boundary conditions on $q$.  A `network' of topological operators for this $\Z_2$ symmetry will be quite simple -- we can either insert the $\Z_2$ generator $\sigma_g$ or not (where not inserting $\sigma_g$ is regarded as inserting the identity operator $\sigma_1$ instead), as shown in Figure~\ref{fig:circws}.  Given any other configuration of insertions of TPOs on such a worldsheet, we would simply fuse all of the operators to end up at either Figure~\ref{circws1} or Figure~\ref{circws2}.

The non-trivial operator insertion on the worldsheet pictured in Figure~\ref{circws2} corresponds to implementing group-twisted boundary conditions on $q$, which is the condition
\be
q(t+2\pi)=g\cdot q(t) = q(t)+\pi.
\ee
Summing over such insertions and normalizing by the order of the group will amount to acting on states with the operator $(1+g)/2$, which projects onto invariant states.  In terms of an operator on the Hilbert space, we can write this as
\be
\label{1dpi+}
\pi^+=\frac{1}{2}(1+g)=\sum_{n\text{ even}}\ket{n}\bra{n}.
\ee

This presentation entirely mimics the usual story for $2d$ CFT orbifolds.  Of course in two dimensions we would recognize that a projection onto group-invariant states was not sufficient to produce a well-behaved CFT and would find ourselves needing to add in twisted states to restore modular invariance.  But in quantum mechanics none of those complications appear, and we are in fact done.

We are allowed to redefine the phases associated with our local operators $\sigma_{g_i}$.  Specifically, we could take $\sigma'_{g_i}=\varphi(g_i)\sigma_{g_i}$.  But the choice of $\varphi$ is not completely arbitrary -- it must be compatible with the fusion relations of the $\sigma$.  We can calculate
\be
\sigma'_{g_1}\otimes\sigma'_{g_2}=\varphi(g_1)\varphi(g_2)\sigma_{g_1}\otimes\sigma_{g_2}=\varphi(g_1)\varphi(g_2)\sigma_{g_1g_2}.
\ee
In order to have $\sigma'_{g_1}\otimes\sigma'_{g_2}=\sigma'_{g_1g_2}$ it must be that
\be
1=\frac{\varphi(g_1)\varphi(g_2)}{\varphi(g_1g_2)},
\ee
which is closure in $H^1(G,U(1))$ or, more familiarly, the requirement that $\varphi$ is a homomorphism from $G$ to $U(1)$.  Choice of $\varphi$ essentially serves as discrete torsion when gauging the symmetry associated to $\sigma_g$.  In the $\Z_2$ example at hand, the gauging that led to (\ref{1dpi+}) corresponds to $\varphi$ trivial.  There is one non-trivial choice of $\varphi$, and this choice leads to the projector
\be
\label{1dpi-}
\pi^-=\frac{1}{2}(1-g)=\sum_{n\text{ odd}}\ket{n}\bra{n}.
\ee
Clearly these projectors sum to the identity operator.  The Hilbert space $\H$ then accordingly breaks into two pieces:
\be
\H=\H^++\H^-.
\ee
In order to phrase this as a sort of decomposition, let us examine the spectrum in each constituent Hilbert space.  Assume that the theory initially has $\theta=0$.  Then the energies in $\H^+$ are
\be
E^+_n=\frac{1}{2}\left(\frac{2n}{R}\right)^2,
\ee
which is the spectrum of the same theory on a circle of radius $R/2$.  The energies of $\H^-$, on the other hand, are given by
\be
E_n^-=\frac{1}{2}\left(\frac{2n-1}{R}\right)^2.
\ee
Comparing the above to (\ref{circen}), we see that $\H^-$ has the spectrum of a theory on a circle of radius $R/2$ but with $\theta=\pi$.  So, if we let $T(R,\theta)$ schematically denote a particle on a circle of radius $R$ with theta term $\theta$, we have the following decomposition of theories:
\be
\label{qmdecomp1}
T(R,0)=T(R/2,0)\oplus T(R/2,\pi).
\ee
Note that a theory breaking into a direct sum with varying theta angles (or choices of discrete torsion) is a common occurrence in decomposition in higher dimensions \cite{Sharpe_Diverse}.  Gauging the 0-form $\Z_2$ symmetry simply selects and projects onto a constituent theory in this decomposition.  Additionally, notice the similarity with the $\Z_2$ shift orbifold of the two-dimensional compact free boson, which also takes a theory with radius $R$ to the same theory at $R/2$.

Similarly, let $S(R)$ denote the quantum mechanical theory of an infinite square well of width $R$.  One can check that
\be
\label{qmdecomp2}
T(R,\pi) = S(\pi R) \oplus S(\pi R).
\ee

As a bit of a tangent, one can leverage (\ref{qmdecomp1}) and (\ref{qmdecomp2}) to play a fun game.  We can write
\begin{align}
T(R,0)&=T(R/2,0)\oplus T(R/2,\pi) \nonumber\\
&=T(R/2,0)\oplus S(\pi R/2) \oplus S(\pi R/2)\\\nonumber
&=T(R/4,0) \oplus T(R/4,\pi) \oplus 2S(\pi R/2)...
\end{align}
to arrive at
\be
T(R,0)=T(0,0)\oplus 2\bigoplus_{n=1}^\infty S\left(\frac{\pi R}{2^n}\right),
\ee
in which the modes of the circle have undergone a sort of Fourier decomposition (double meaning intended) into an infinite sum of square wells with ever-increasing ground state energies, plus some sort of instanton describing the circle zero mode, which takes the form (formally) of a sigma model of a point. 

\subsection{The Quantum Symmetry}

\begin{figure}[h]
	\begin{subfigure}{0.5\textwidth}
	\centering
	\begin{tikzpicture}
	\draw[thick,dashed] (0,0) [partial ellipse=0:360:1cm and 1cm];
	\end{tikzpicture}
	\caption{}
	\label{circwsline1}
	\end{subfigure}
	\begin{subfigure}{0.5\textwidth}
	\centering
	\begin{tikzpicture}
	\draw[thick] (0,0) [partial ellipse=0:360:1cm and 1cm];
	\end{tikzpicture}
	\caption{}
	\label{circwsline2}
	\end{subfigure}
\caption{Two ways of wrapping TDLs corresponding to a $-1$-form $\Z_2$ symmetry on a 1$d$ worldsheet.}
\label{fig:circwsline}
\end{figure}
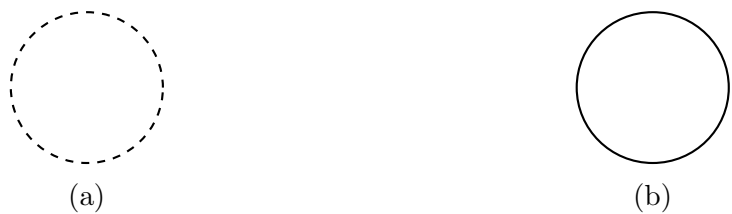

Let's say we have gauged our theory $T(R,0)$ by projecting onto the $\H^+$ Hilbert space to arrive at $T(R/2,0)$.  Does there, as the general claim goes, exist a dual quantum symmetry which we can gauge to return to our original theory?  Based on the discussion so far, such a symmetry would be a $-1$-form symmetry implemented by line operators.  We can construct such a picture by considering a pair of line operators associated to the theta parameter.  The associated compactified worldlines are shown in Figure~\ref{fig:circwsline}.  The dashed line of Figure~\ref{circwsline1} indicates that the identity line operator has been inserted, in contrast to the solid line of Figure~\ref{circwsline2} which is meant to indicate insertion of the non-trivial line operator.

The action of the non-trivial TDL will be to shift the theta parameter of the theory by $\pi$.  Gauging the $-1$-form symmetry is implemented by summing over insertions of such worldline TDLs, producing
\be
T(R/2,0) \oplus T(R/2,\pi).
\ee
By (\ref{qmdecomp1}) this indeed recovers $T(R,0)$, undoing the gauging.  If we had projected onto $\H^-$ instead, an identical procedure would apply, except that now the identity TDL insertion would correspond to $T(R/2,\pi)$ and the non-trivial insertion would produce $T(R/2,0)$.

If we had instead started from $T(R,\pi)$, (\ref{qmdecomp2}) tells us that there is only one distinct choice of gauging, as both universes in the decomposition are identical.  The corresponding $-1$-form symmetry is then trivially-acting, in the sense that whichever of the two TDLs we choose to wrap the worldline, we should obtain $S(\pi R)$.  Gauging the dual symmetry then simply corresponds to adding these two copies together.

While these operations on $1d$ theories may seem quaint, they provide a simple introduction to the notion of a $-1$-form symmetry and provide an important consistency check of the claims made in section~\ref{sec:intro}.  In the remainder of the paper we will see that higher dimensions admit a similar story, made more exciting by the fact that we have additional types of symmetry which can mix in.

\section{Preliminaries in Two Dimensions}
\label{sec:prelim}

Now we advance from quantum mechanics back to full quantum field theory in $(1+1)d$.  Before considering how $-1$-form symmetries appear in this context, we need to lay some groundwork.  This will involve reviewing known material regarding 0-form and 1-form symmetries, their behavior and associated manipulations in two dimensions.  While most of the quantiative contents of this section are known, we hope to incorporate some new perspectives on the material along the way.

\subsection{Decomposition and Trivially-Acting Symmetries}
\label{sec:decompreview}

In section~\ref{sec:qm} we saw examples of a symmetry causing a quantum mechanical system to break up as a direct sum of theories.  Does something analogous happen in two dimensions?  The answer is yes, but we need to be careful.  When we speak of a quantum mechanical system with an `ordinary' (0-form) symmetry, the correct comparison in $2d$ would be a system with 1-form symmetry -- both of these are $(d-1)$-form symmetries, and both are associated to local topological operators.

Thus we should expect a $2d$ system with a 1-form symmetry to decompose.  In fact, in 2d conformal field theory (CFT) there is a very simple way to see this.  Given a non-trivial spectrum of TPOs with group-like fusion relations, we can rearrange these operators into linear combinations that act as projectors (just as we did in (\ref{1dpi+}) and (\ref{1dpi-})).  By the state-operator correspondence, these weight zero operators should map to states in the system.  These states are exactly the vacua of the universes in the decomposition.

There is a convenient way to construct such a system as an orbifold \cite{TopOps}.  If a theory has a trivially-acting 0-form symmetry, there exist TDLs corresponding to that symmetry.  TDLs differing by a trivially-acting symmetry can form two-way junctions, and the local operators at those junctions furnish the theory with a non-trivial spectrum of local topological operators.  Upon gauging such a symmetry, the junction operators form a subset of twist fields in the gauged theory, and are `freed' from the lines to which they were previously bound.  This allows them to form projection operators and hence vacua in a decomposition.

The upshot of this is that an orbifold theory in which the orbifold group acted non-effectively (i.e.~at least some subgroup of it was trivially-acting) will in general have a non-trivial spectrum of local topological operators, and therefore will be equivalent to a direct sum of theories.  Note, however, that a disjoint sum of theories could have multiple presentations as a gauged theory.  For instance, take a direct sum of four copies of a theory.  We could obtain such a sum by taking a single copy of the theory in question and gauging a trivially-acting $\Z_4$ symmetry, or by gauging a trivially-acting $\Z_2\times\Z_2$ symmetry in the same theory.\footnote{One could attempt to use the associated quantum symmetry as a means of distinguishing these two cases.  One of these gaugings should have a $\hat{\Z}_4$ symmetry, while the other should have a $\hat{\Z}_2\times\hat{\Z}_2$.  However, both of these symmetries should manifest as subgroups of the $S_4$ exchange symmetry enjoyed by the four copies.  In general we could also consider adding in effectively-acting symmetries of the theory and non-trivial mixings between exchange of copies and the effective symmetries of each single copy.  Despite such considerations, all of these approaches produce the same local theory.}  As we develop additional machinery, we will have more to say about the notion of `supplementing' a theory with a trivially-acting symmetry and related interpretational issues.\\

The above description of trivially-acting symmetries as a mix of topological line and point operators can be given a useful shorthand.  In general, we would like to use the notation
\be
\label{rtimes}
A_{[p]} . B_{[q]}
\ee
for a mix of a $p$-form symmetry with a $q$-form symmetry.  How should we understand (\ref{rtimes}) mathematically?  One way would be to define $A_{[p]}. B_{[q]}$ by the existence of a short exact sequence
\be
1\lrr A_{[p]}\lrr A_{[p]}. B_{[q]}\lrr B_{[q]}\lrr 1,
\ee
which \cite{Tachikawa} suggests should be classified by 
\be
\label{mixedextclass}
H^{p+2}(\mathcal{K}(B,q+1),A),
\ee
with $\mathcal{K}(G,m)$ the $m$th Eilenberg-Mac Lane space associated to $G$.  For $m>1$, such a space is only defined for abelian $G$, but we would like to allow our symmetries to be as general as possible (for example, non-abelian or non-group-like).  Attempting to find a sufficiently broad formal definition would get us farther into the weeds of higher category theory than is necessary for this paper, so we would settle for a more pedestrian definition (though we will still utilize (\ref{mixedextclass}) in situations where it applies).

It is clear from a physical perspective what $A_{[p]}\times B_{[q]}$ should mean -- such an expression should describe a non-interacting mix of $p$-form and $q$-form symmetries in a theory.  That is, a theory with such a symmetry would have codimension $p+1$ and codimension $q+1$ topological operators describing independent $p$-form and $q$-form symmetries.  Accordingly, we will take $A_{[p]}. B_{[q]}$ to mean that a theory has codimension $p+1$ and codimension $q+1$ topological operators which mix in some non-trivial fashion, such that the symmetry $B_{[q]}$ does not necessarily constitute a subsymmetry of the system -- that is, (even in the absence of any form of gauge anomaly) we may not be able to gauge $B_{[q]}$ by itself.  The obstruction to $B_{[q]}$ constituting a standalone symmetry would serve the role of the extension class.  Additionally, in analogue to an ordinary non-central extension, there may be some action of the operators associated to $B_{[q]}$ on those associated with $A_{[p]}$ or vice versa.  When $p=q=0$, such a description matches the notation of \cite{ATLAS} for group extensions which are not necessarily split and not necessarily central.

In this language, we would say that a $2d$ theory with trivially-acting symmetry $K$ has the symmetry $K_{[0]}. K_{[1]}$, where the 1-form symmetry is given by the TPOs (the twist fields of weight zero) living at the junctions between TDLs.  As the TPOs are constrained to live at junctions, they cannot be inserted into correlation functions independent of the TDLs which generate the 0-form symmetry.  For abelian $K$, we can actually leverage (\ref{mixedextclass}) to explicitly calculate the extension class.  Using the fact that Eilenberg-Mac Lane spaces are defined by the condition $\pi_n(\mathcal{K}(G,n))=G$, we have
\begin{multline}
\label{emcalc}
H^2(\mathcal{K}(K,2),K)=\text{Hom}(H_2(\mathcal{K}(K,2),\Z),K)\\
=\text{Hom}(\pi_2(\mathcal{K}(K,2)),K)=\text{Hom}(K,K).
\end{multline}

More generally, if the theory also includes an effective zero-form symmetry $G_{[0]}$, the total symmetry would be $(K_{[0]}. G_{[0]}). K_{[1]}$.  Note that there is a natural action of the line operators specifying the 0-form part of this symmetry on the local operators which generate the 1-form part: if we write the total 0-form symmetry as $\Gamma_{[0]}=K_{[0]}. G_{[0]}$, since $K$ is a normal subgroup of $\Gamma$ a TDL labeled by $\gamma$ can act on a TPO labeled by $K$ as $\gamma\cdot\sigma_k=\sigma_{\gamma k\gamma^{-1}}$.\\

We can see an example of such a setup that does not feature decomposition by briefly moving to (2+1) dimensions to examine $\Z_2$ gauge theory, which is a theory with a gauged, trivially-acting $\Z_2$ 0-form symmetry.  Based on the discussion above, we would say that the global symmetry of the ungauged theory should be $(\Z_2)_{[0]}. (\Z_2)_{[1]}$, where the operators corresponding to the 1-form symmetry are constrained to live at junctions of the operators which generate the 0-form symmetry.  The extension class would be valued in $H^2(\mathcal{K}(\Z_2,2),\Z_2)=\text{Hom}(\Z_2,\Z_2)=\Z_2$, and would in particular be the isomorphism from $\Z_2$ to $\Z_2$.

The $\Z_2$ gauge theory is then obtained by gauging $(\Z_2)_{[0]}$.  From the discussion in \cite{Tachikawa}, we would expect the resulting theory to have symmetry $(\hat{\Z}_2)_{[1]}\times(\Z_2)_{[1]}$, and the non-trivial extension class from the ungauged theory should determine a non-trivial mixed anomaly in the gauged theory.  This matches exactly with our expectations -- $3d$ $\Z_2$ gauge theory is known to have two $\Z_2$ 1-form symmetries.  To match with the usual terminology, $(\hat{\Z}_2)_{[1]}$ would be called electric and $(\Z_2)_{[1]}$ would be called magnetic.  And indeed, these electric and magnetic symmetries are known to have a mixed anomaly, entirely in line with the above formulation.  For a closely related analysis, see \cite[Section 3.1]{Tachikawa}.

\subsection{Gauging Symmetries in Two Dimensions}
\label{sec:gaugeone}

We begin with a general observation.  Gauging any symmetry can be thought of as coupling to a background gauge field for that symmetry and summing over configurations.  For QFT on a manifold $M$, a background gauge field for a $p$-form symmetry $G_{[p]}$ takes values in $H^{p+1}(M,G)$.  Then for an ordinary (0-form) symmetry on a 2-torus, we have
\be
H^1(S^1\times S^1,G)=G\otimes G,
\ee
which is to say we get two copies of the group $G$.  The $T^2$ partition function of an orbifold should then be expressible as a sum over two copies of the group, which is a familiar result.  Additionally, we have
\be
H^2(S^1\times S^1,G)=G,
\ee
which as we will see in this section tells us that gauging by a 1-form symmetry should involve a single sum over the group.  For a $-1$-form symmetry, then, we would appeal to the result
\be
\label{h0t2}
H^0(S^1\times S^1,G)=G
\ee
to set our expectation that the genus one partition funcion for gauging a $-1$-form symmetry $G$ should also be expressible as a single sum over $G$.  Note that all of these calculations are in harmony with the $1d$ results, as
\be
H^0(S^1,G)=H^1(S^1,G)=G
\ee
so in quantum mechanics on a compact worldline, one would expect gauging both 0-form and 1-form symmetries to involve a single sum over the group, which is precisely what our example in section~\ref{sec:qm} produced.\\

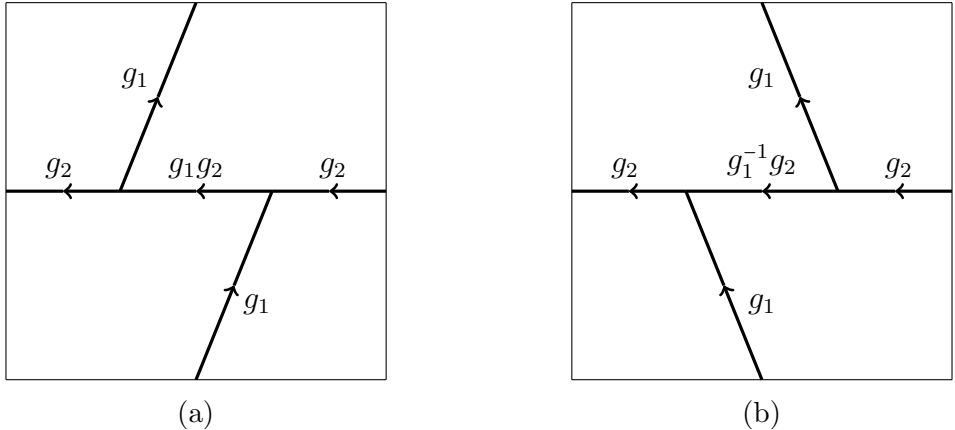
\begin{figure}[h]
	\begin{subfigure}{0.5\textwidth}
	\centering
	\begin{tikzpicture}
	\draw[thin] (0,0)--(5,0);
	\draw[thin] (5,0)--(5,5);
	\draw[thin] (0,0)--(0,5);
	\draw[thin] (0,5)--(5,5);
	\draw[very thick,->] (2.5,0)--(3,1.25);
	\draw[very thick] (3,1.25)--(3.5,2.5);
	\draw[very thick,->] (1.5,2.5)--(2,3.75);
	\draw[very thick] (2,3.75)--(2.5,5);
	\draw[very thick,->] (5,2.5)--(4.25,2.5);
	\draw[very thick] (4.25,2.5)--(3.5,2.5);
	\draw[very thick,->] (3.5,2.5)--(2.5,2.5);
	\draw[very thick] (2.5,2.5)--(1.5,2.5);
	\draw[very thick,->] (1.5,2.5)--(0.75,2.5);
	\draw[very thick] (0.75,2.5)--(0,2.5);
	\node at (1.7,4) {$g_1$};
	\node at (3.3,1) {$g_1$};
	\node at (0.7,2.8) {$g_2$};
	\node at (4.3,2.8) {$g_2$};
	\node at (2.5,2.8) {$g_1g_2$};
	\end{tikzpicture}
	\caption{}
	\label{fig:orb_pf_sec1}
	\end{subfigure}
	\begin{subfigure}{0.5\textwidth}
	\centering
	\begin{tikzpicture}
	\draw[thin] (0,0)--(5,0);
	\draw[thin] (5,0)--(5,5);
	\draw[thin] (0,0)--(0,5);
	\draw[thin] (0,5)--(5,5);
	\draw[very thick,->] (2.5,0)--(2,1.25);
	\draw[very thick] (2,1.25)--(1.5,2.5);
	\draw[very thick,->] (3.5,2.5)--(3,3.75);
	\draw[very thick] (3,3.75)--(2.5,5);
	\draw[very thick,->] (5,2.5)--(4.25,2.5);
	\draw[very thick] (4.25,2.5)--(3.5,2.5);
	\draw[very thick,->] (3.5,2.5)--(2.5,2.5);
	\draw[very thick] (2.5,2.5)--(1.5,2.5);
	\draw[very thick,->] (1.5,2.5)--(0.75,2.5);
	\draw[very thick] (0.75,2.5)--(0,2.5);
	\node at (2.5,4) {$g_1$};
	\node at (2.5,1) {$g_1$};
	\node at (0.7,2.8) {$g_2$};
	\node at (4.3,2.8) {$g_2$};
	\node at (2.5,2.9) {$g_1^{-1}g_2$};
	\end{tikzpicture}
	\caption{}
	\label{fig:orb_pf_sec2}
	\end{subfigure}
	\caption{The $(g_1,g_2)$ sector of an orbifold partition function, with a four-way junction resolved into three-way junctions in two different ways.}
	\label{fig:orb_pf_sec}
\end{figure}

While the notion of gauging a 0-form symmetry (orbifolding) in two dimensions is likely quite familiar, we review it here for completeness.  The presentation will be along the lines of \cite{ChangLinShaoYin}.  We will begin by describing the effect of gauging on the genus one partition function -- an analogous process applies on higher genus surfaces.  Picturing the torus as its fundamental domain in $\C$, Figure~\ref{fig:orb_pf_sec} shows the effect of wrapping the homotopy cycles of the surface with TDLs labeled by $g_1$ and $g_2$.  The fields that are well-defined on such a decorated surface are ones that close up to group transformations.  In the path integral formulation, we would implement the gauging by summing over fields satisfying such group-twisted boundary conditions \cite{Ginsparg}:
\be
\phi(z+1,\bar{z}+1)=g_1\cdot\phi(z,\bar{z})\hspace{0.5cm}\phi(z+\tau,\bar{z}+\bar{\tau})=g_2\cdot\phi(z,\bar{z})
\ee
where $\tau$ is the complex structure constant for a torus given by a lattice with basis vectors $(1,\tau)\in\C$.  Alternatively, we can view such a configuration as a discrete gauge bundle in which the TDLs play the role of transition functions \cite{Tachikawa_Lectures}.

In order for this bundle to be constructed in a gauge-invariant manner, we should be able to join networks of lines at topological junctions in any fashion consistent with the group laws.  For instance, Figures~\ref{fig:orb_pf_sec1} and \ref{fig:orb_pf_sec2} depict the same four lines joined at a pair of three-way junctions in two different ways.  In principle, swapping two such resolutions of a four-way junction where $g_1$, $g_2$ and $g_3$ meet to form $g_1g_2g_3$ could introduce a phase $\omega(g_1,g_2,g_3)$.  Consistency when resolving a five-way junction further fixes $\omega$ up to its class in $H^3(G,U(1))$, leading to the well-known cohomological classification of gauge anomalies of $G$.  For the remainder of this paper we will assume that any 0-form symmetries present are free of such gauge anomalies, i.e.~the associated class in $H^3(G,U(1))$ is trivial.

In either picture, it is clear that $g_1$ and $g_2$ must commute.  As boundary conditions, $\phi(z+1+\tau,\bar{z}+1+\bar{\tau})$ would be ambiguous if $[g_1,g_2]\neq 1$.  In terms of the TDL picture, the problem arises when we demand consistency with the group law.  For instance, if we invert the leftmost $g_1$ and $g_2$ lines of Figure~\ref{fig:orb_pf_sec1} such that all lines are incoming to the four-way junction, it should be that the product of the lines is the identity in $G$.  However, for this product (say, starting at $g_1$ and going counter-clockwise) we find $g_1g_2g_1^{-1}g_2^{-1}=[g_1,g_2]$.  Therefore the configuration drawn in Figure~\ref{fig:orb_pf_sec1} is not consistent for $g_1$ and $g_2$ not commuting.\footnote{An alternative way to see this in the TDL picture is to demand that the rightmost junction of Figure~\ref{fig:orb_pf_sec1} is formed as drawn, i.e.~with $g_1$ and $g_2$ incoming and fusing into $g_1g_2$.  Then in order to, say, split off an outgoing $g_2$ as drawn, the line labeled by $g_1$ would need to be labeled instead by $g_2^{-1}g_1g_2$.  Of course when $[g_1,g_2]=1$ these are the same thing and the diagram makes sense as drawn, but when the two do not commute we would be left with a $g_2^{-1}g_1g_2$ line which we would have no hope of consistently joining to the $g_1$ line.}  Summing over consistent configurations and dividing by the order of the gauge group, the partition function of the $G$ orbifold is obtained as
\be
\frac{1}{|G|}\sum_{\substack{g_1,g_2\in G \\ [g_1,g_2]=1}}Z_{g_1,g_2}.
\ee

\begin{figure}[h]
	\begin{subfigure}{0.5\textwidth}
	\centering
	\begin{tikzpicture}
	\draw[very thick,->] (-2,2) -- (-1.5,1);
	\draw[very thick] (-1.5,1) -- (-1,0);
	\node at (-1,1.5) {$g_1$};
	\draw[very thick,->] (-2,-2) -- (-1.5,-1);
	\draw[very thick] (-1.5,-1) -- (-1,0);
	\node at (-1,-1.5) {$g_2$};
	\draw[very thick,->] (-1,0) -- (0,0);
	\draw[very thick] (0,0) -- (1,0);
	\node at (0,0.5) {$g_1g_2$};
	\draw[very thick,->] (1,0) -- (1.5,1);
	\draw[very thick] (1.5,1) -- (2,2);
	\node at (1,1.5) {$g_1g_2g_3$};
	\draw[very thick](1.5,-1) -- (1,0);
	\draw[very thick,->] (2,-2) -- (1.5,-1);
	\node at (1,-1.5) {$g_3$};
	\filldraw[black] (-1,0) circle (2pt);
	\node at (-2.25,0) {$\varepsilon(g_1,g_2)$};
	\filldraw[black] (1,0) circle (2pt);
	\node at (2.5,0) {$\varepsilon(g_1g_2,g_3)$};
	\end{tikzpicture}
	\caption{}
	\label{dt_swap1}
	\end{subfigure}
	\begin{subfigure}{0.5\textwidth}
	\centering
	\begin{tikzpicture}
	\draw[very thick,->] (-1,2) -- (-0.5,1.5);
	\draw[very thick] (-0.5,1.5) -- (0,1);
	\node at (-1,1.5) {$g_1$};
	\draw[very thick,->] (0,1) -- (0.5,1.5);
	\draw[very thick] (0.5,1.5) -- (1,2);
	\node at (1.4,1.5) {$g_1g_2g_3$};
	\draw[very thick,->] (0,-1) -- (0,0);
	\draw[very thick] (0,0) -- (0,1);
	\node at (-0.5,0) {$g_2g_3$};
	\draw[very thick,->] (-1,-2) -- (-0.5,-1.5);
	\draw[very thick] (-0.5,-1.5) -- (0,-1);
	\node at (-1,-1.5) {$g_2$};
	\draw[very thick,->] (1,-2) -- (0.5,-1.5);
	\draw[very thick] (0.5,-1.5) -- (0,-1);
	\node at (1,-1.5) {$g_3$};
	\filldraw[black] (0,1) circle (2pt);
	\node at (1.25,0.75) {$\varepsilon(g_1,g_2g_3)$};
	\filldraw[black] (0,-1) circle (2pt);
	\node at (1,-0.75) {$\varepsilon(g_2,g_3)$};
	\end{tikzpicture}
	\caption{}
	\label{dt_swap2}
	\end{subfigure}
	\caption{A four-way junction decomposed into three-way junctions with phases.}
	\label{fig:dt_swap}
\end{figure}
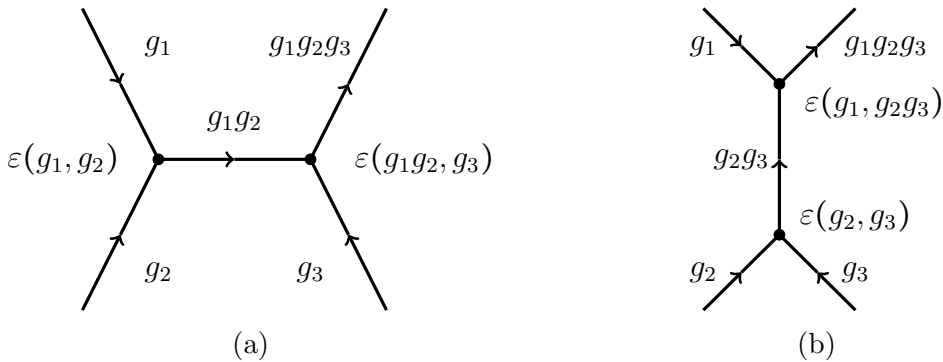

At junctions where TDLs meet, there can sit local operators.\footnote{The following discussion assumes that the group $G$ acts effectively on the states of our theory.  As was mentioned in the previous subsection, the spectrum of local topological operators at junctions can be richer in the presence of a trivially-acting symmetry.}  The junction Hilbert space of weight zero operators for any configuration of TDLs consistent with the group laws is isomorphic to $\C$ -- that is, at a junction where $g_1$ and $g_2$ meet to form $g_1g_2$, we could have a phase $\varepsilon(g_1,g_2)$.  (For configurations inconsistent with the group law, we could say that the junction Hilbert space contains no weight zero operators.  This is another way of saying that such configurations must vanish.)  Of course there will be consistency conditions on this phase.  In particular, as mentioned above, in the absence of gauge anomalies we should be able to swap four-way junctions freely.  Figures~\ref{dt_swap1} and \ref{dt_swap2} show the result of such a swap, along with the phases that sit at each three-way junction.  Since the total phase should be the same when swapping, we find the condition
\be
\varepsilon(g_1,g_2)\varepsilon(g_1g_2,g_3)=\varepsilon(g_2,g_3)\varepsilon(g_1,g_2g_3)
\ee
which tells us that $\varepsilon$ is a 2-cocycle on $G$.

This assignment of phases to junctions is not unique.  Much like in section~\ref{sec:qmgauging}, we can redefine the contribution of each TDL to such a junction.  We do so by choosing a map $\lambda(g):G\to U(1)$ and shifting the contribution of an incoming $g$ line by $\lambda(g)$.  This means that $\varepsilon(g_1,g_2)$ can be freely shifted by $\lambda(g_1)\lambda(g_2)\lambda^{-1}(g_1g_2)$, thus $\varepsilon$ is determined up to its class in $H^2(G,U(1))$.  The effect of a non-trivial $\varepsilon$ is to introduce phases\footnote{In this case, the phase assignment conventions are usually chosen such that the phase assigned to $Z_{g_1,g_2}$ is $\varepsilon(g_1,g_2)\varepsilon^{-1}(g_2,g_1)$.} into the genus one (and higher) partition function(s); this phenomenon is better known as discrete torsion.\\

The gauging of 1-form symmetries in two dimensions may be less familiar, but the process will be essentially the same as in section~\ref{sec:qmgauging}.  Having a 1-form symmetry described by an abelian group $K$ means that there exist topological point operators which fuse according to the group rules in $K$.  In order to calculate the effect on the genus one partition function, we sum over topologically distinct insertions of these TPOs.  As these are local operators, they are insensitive to the topology of the (connected) Riemann surface that  serves as our $2d$ worldsheet, but for ease of comparison we will continue to use a torus.  Accordingly, let $Z_{\sigma_k}$ denote the torus correlation function of the TPO labeled by $k$.  Any possible configuration of insertions of TPOs $\sigma_{k_i}$ can be reduced to $Z_{\sigma_{\prod_ik_i}}$ by fusion, so we only need to sum over a single copy of $K$.  The torus partition function then takes the form
\be
\label{1gaugepf}
\frac{1}{|K|}\sum_{k\in K}Z_{\sigma_k}.
\ee
Now we note that this expression could be interpreted as the insertion of a single composite local operator
\be
\pi_1=\frac{1}{|K|}\sum_{k\in K}\sigma_k.
\ee\
Recalling that the $\sigma_k$ fuse as $\sigma_{k_1}\otimes\sigma_{k_2}=\sigma_{k_1k_2}$, one readily sees that $\pi_1\otimes\pi_1=\pi_1$.  So gauging our 1-form symmetry is equivalent to inserting the projector $\pi_1$ onto the worldsheet (and of course inserting it once is the same as inserting it any number of times).  Thus, we can regard gauging the 1-form symmetry as projection onto the universe whose vacuum is given by the state corresponding to the operator $\pi_1$ \cite{Sharpe_UndoingDecomposition}.

Of course, this choice of gauging was not unique.  Once again we should expect the possibility of a choice of discrete torsion, which here should be classified by $H^2(\mathcal{K}(K,2),U(1))$.  Happily, the calculation (\ref{emcalc}) can be repeated to tell us that discrete torsion in this gauging is given by a choice of $\text{Hom}(K,U(1))=\hat{K}$, the Pontryagin dual to $K$.  That is, given a $\varphi:K\to U(1)$ corresponding to an element $\hat{k}$ of $\hat{K}$, we have 
\be
\label{khatproj}
\pi_{\hat{k}}=\frac{1}{|K|}\sum_{k\in K}\varphi(k)\sigma_k.
\ee
Gaugings of $K_{[1]}$ are then quite simply given by insertion of a projector $\pi_{\hat{k}}$, and there are $|\hat{K}|=|K|$ such choices.  The effect is to project onto one of the $|K|$ universes in the decomposition associted with $K_{[1]}$.

In the previous subsection \ref{sec:decompreview}, we noted that a theory exhibiting decomposition may have multiple realizations as an orbifold by non-effective symmetries.  The above description of gauging should shed additional light on this matter -- ultimately, the group structure of $K$ is irrelevant to the outcome.  If we have a theory that decomposes into $N$ components, there will be $N$ vacua and hence $N$ projection operators.  Gauging the 1-form symmetry simply involves selecting one of those projectors and flooding the worldsheet with it, which projects onto its associated universe.  We could have obtained this decomposition in any number of ways as an orbifold -- the twist fields coming from that orbifold could have rather exotic fusion relations (see \cite{Sharpe_Noninvertible} for some examples), but at the end of the day we will be able to recombine them into $N$ projection operators and forget about their origin.

\subsection{Mixed Gauging}

Section~\ref{sec:decompreview} claimed that a theory with effective 0-form symmetry $G$ and trivially-acting symmetry $K$ has the symmetry $(K_{[0]}. G_{[0]}). K_{[1]}$.  Now that we have reviewed the separate gauging of 0-form and 1-form symmetries, could we make sense of gauging such a mix of symmetries?  That this should be possible in general would follow from a stronger notion of orbifold composibility than the one reviewed in section~\ref{sec:intro} -- we would want it to be the case that gauging a theory $T$ by a $p$-form symmetry $A_{[p]}$ and then gauging the resulting theory by a $q$-form symmetry $B_{[q]}$ should be the same as gauging $T$ by $A_{[p]}.B_{[q]}$.  We will construct such a gauging explicitly for the theory in question, for which we expect that
\be
T/[(K_{[0]}. G_{[0]}). K_{[1]}] = [T/(K_{[0]}. G_{[0]})]/K_{[1]}.
\ee
The only subtlety in carrying out this prescription will be that, when inserting the projector associated with gauging the 1-form symmetry, we must identify its preimage in the original theory $T$.

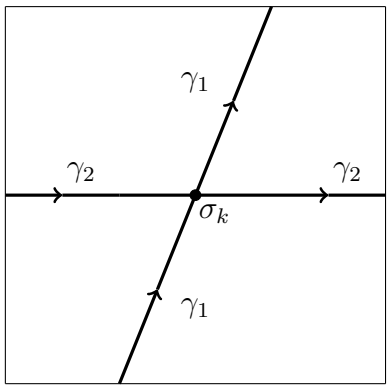
\begin{figure}
	\centering
	\begin{tikzpicture}
	\draw[thin] (0,0)--(5,0);
	\draw[thin] (5,0)--(5,5);
	\draw[thin] (0,0)--(0,5);
	\draw[thin] (0,5)--(5,5);
	\draw[very thick,->] (1.5,0)--(2,1.25);
	\draw[very thick] (2,1.25)--(2.5,2.5);
	\draw[very thick,->] (2.5,2.5)--(3,3.75);
	\draw[very thick] (3,3.75)--(3.5,5);
	\draw[very thick] (4.25,2.5)--(5,2.5);
	\draw[very thick,->] (2.5,2.5)--(4.25,2.5);
	\draw[very thick] (2.5,2.5)--(1.5,2.5);
	\draw[very thick] (0.75,2.5)--(1.5,2.5);
	\draw[very thick,->] (0,2.5)--(0.75,2.5);
	\node at (2.5,4) {$\gamma_1$};
	\node at (2.5,1) {$\gamma_1$};
	\node at (1,2.8) {$\gamma_2$};
	\node at (4.5,2.8) {$\gamma_2$};
	\filldraw[black] (2.5,2.5) circle (2pt);
	\node at (2.75,2.25) {$\sigma_k$};
	\end{tikzpicture}
	\caption{The visual representation of $Z_{\gamma_1,\gamma_2,\sigma_k}$.}
	\label{fig:mg}
\end{figure}

Specifically, let the total 0-form symmetry of $T$ be $\Gamma_{[0]}=K_{[0]}. G_{[0]}$.  The torus partition function of the orbifold $T/\Gamma_{[0]}$ is
\be
\label{mixg1}
\bar{Z}=\frac{1}{|\Gamma|}\sum_{\substack{\gamma_1,\gamma_2\in \Gamma \\ [\gamma_1,\gamma_2]=1}}Z_{\gamma_1,\gamma_2}.
\ee
Now in order to gauge $T/\Gamma_{[0]}$ by its 1-form symmetry $K_{[1]}$, we choose a projector labeled by $\varphi(k)$ corresponding to an element $\hat{k}$ of $\hat{K}$ and project onto its associated universe.  For ease, let us take the trivial choice of $\varphi$.  By (\ref{1gaugepf}) and (\ref{khatproj}), the resulting partition function takes the form
\be
\label{mixg2}
\frac{1}{|K|}\sum_{k\in K}\bar{Z}_{\sigma_k}.
\ee
In order to combine (\ref{mixg1}) and (\ref{mixg2}), we note that in the ungauged theory $T$, the operators $\sigma_k$ are confined to live at intersections of TDLs.  Fortunately we have such an intersection, between the TDLs $\gamma_1$ and $\gamma_2$.  This motivates us to define the object $Z_{\gamma_1,\gamma_2,\sigma_k}$ in the theory $T$ as the torus wrapped with $\gamma_1$ and $\gamma_2$ with $\sigma_k$ placed at their intersection.  This configuration is shown in Figure~\ref{fig:mg}.

Then, the doubly-gauged partition function of $(T/\Gamma_{[0]})/K_{[1]}$ should be expressible as
\be
\label{mgpf}
\frac{1}{|K|}\frac{1}{|\Gamma|}\sum_{\substack{\gamma_1,\gamma_2\in \Gamma \\ k\in K \\ [\gamma_1,\gamma_2]=k}}Z_{\gamma_1,\gamma_2,\sigma_k},
\ee
and this would be the partition function for $T/(\Gamma_{[0]}. K_{[1]})$.  Note that the commutation constraint on $\gamma_1$ and $\gamma_2$ has been broadened.  In the case of effectively-acting symmetries, only the identity TPO $\sigma_1$ could sit at an intersection of $\gamma_1$ and $\gamma_2$.  We would then only be able to wrap the homotopy cycles of the torus with such lines when $[\gamma_1,\gamma_2]=1$, as any wrapping by TDLs that did not meet this condition would violate the group law (so perhaps the better way to phrase it is that those configurations were always meant to be included in the sum, but normally would vanish).  In the case at hand, however, the presence of $\sigma_k$ allows configurations where $\gamma_1$ and $\gamma_2$ do not commute to remain consistent with the group laws.\\

Mixed gauging allows us to solve a potentially nagging issue related to trivially-acting symmetries.  Suppose we begin with a theory $T$ having an effectively-acting 0-form symmetry $G$.  Our attitude so far has been that we can freely add to $T$ a trivially-acting $K$ symmetry for use in calculations, and that the total 0-form symmetry $\Gamma$ could even be a non-trivial mix of $G$ and $K$.  In such a case where $\Gamma$ is a non-trivial extension of $G$ by $K$, it would seem that $G$ ceases to be a proper subsymmetry of the system, in the sense that we would have no hope of gauging $G$ by itself.  Indeed, if the symmetries under consideration were all effectively-acting, we would conclude that $G$ is in general not a subgroup of $\Gamma= K. G$, and we would have no hope of sensibly constructing the gauged theory $T/G$.\footnote{Of course we could first gauge $K$, and then the resulting theory would have $G=\Gamma/K$ symmetry.  Composibility gives us $(T/K)/G=T/\Gamma$.  But this is not the same as $G$ being a symmetry of $T$.}

However, the case under consideration is one in which $K$ acts trivially, so following section~\ref{sec:decompreview} we would say that adding in such a symmetry gives us total symmetry $(K_{[0]}. G_{[0]}). K_{[1]}$.  Now consider applying decomposition to (\ref{mgpf}).  Specifically, let $\rho:\Gamma\to G$ be the projection arising from the short exact sequence $1\to K\to\Gamma\to G\to 1$.  As $K$ acts trivially, decomposition should map each sector $Z_{\gamma_1,\gamma_2,\sigma_k}$ to $Z_{\rho(\gamma_1),\rho(\gamma_2),\sigma_1}$.  Consistency with the group laws implies that we must have
\be
\label{kcommcond}
[\gamma_1,\gamma_2]=k \in K,
\ee
which can only be satisfied for
\be
\label{gcommcond}
[\rho(\gamma_1),\rho(\gamma_2)]=1\in G.
\ee
Using (\ref{kcommcond}) to pick out a single value in the sum over $K$, (\ref{mgpf}) becomes
\be
\frac{1}{|K|}\frac{1}{|\Gamma|}\sum_{\substack{\gamma_1,\gamma_2\in \Gamma \\ [\rho(\gamma_1),\rho(\gamma_2)]=1}}Z_{\rho(\gamma_1),\rho(\gamma_2),\sigma_1},
\ee
Now we write $\rho(\gamma_1)=g_1$, $\rho(\gamma_2)=g_2$ and reinterpret the above in terms of sectors $Z_{g_1,g_2}$ of a $G$ orbifold.  Keeping in mind that $|\Gamma|=|K||G|$, we can rewrite the double sum over $\Gamma$ as a double sum over $G$ that's overcounted by $|K|^2$ (as the $K$ part of the sum is now unconstrained), reducing (\ref{mgpf}) to
\be
\frac{1}{|G|}\sum_{\substack{g_1,g_2\in G \\ [g_1,g_2]=1}}Z_{g_1,g_2},
\ee
which is prcisely the partition function for a $G$ orbifold.  We can summarize this calculation as
\be
\label{trivsym_mg}
T/[(K_{[0]}. G_{[0]}). K_{[1]}]=T/G_{[0]}
\ee

In one sense this result is not at all surprising.  It is a basic result of decomposition that taking a theory with symmetry $G$, adding a trivially-acting abelian symmetry $K$, and orbifolding by the resulting symmetry produces a direct sum of $|K|$ copies of the $G$ orbifold.  Subsequent gauging of the 1-form symmetry in the gauged theory clearly has to produce a copy of the $G$ orbifold.  But from the point of view discussed above, this is quite a surprising result.  We began with a theory having $G_{[0]}$ symmetry, supposed that its total 0-form symmetry no longer included $G_{[0]}$ as a subsymmetry, and yet still managed to construct a gauging of that theory by $G_{[0]}$.  To reiterate, this is a property specifically tied to the trivial action of $K$.

\subsection{Quantum Symmetries and Composibility}
\label{sec:qscomp}

To close out the preparatory material, let us consider how quantum symmetries interact with composition of orbifolds (similar considerations appear from a different perspective in \cite{GaiottoKulp}).  Let us begin with a two-dimensional theory $T$ having an abelian symmetry 0-form symmetry $G$.  The gauged theory $T/G$ has an effectively-acting symmetry $\hat{G}$, the elements of which act on a field in the $g$-twisted sector as multiplication by the phase $\chi_{\hat{g}}(g)$.  Gauging this quantum symmetry returns us to the original theory, so $(T/G)/\hat{G}=T$.

Now we would like to invoke composibility of orbifold operations to argue that there should be some symmetry of $T$, call it $\io_G(\hat{G})$, such that\footnote{The assertion that the resulting symmetry takes the form of a direct product with $G$ is slightly stronger than the general statement of composibility, but in this case it will turn out to be true.}
\be
\label{qscomp}
T/(G\times\io_G(\hat{G}))=T.
\ee
$\io_G(\hat{G})$ must exist for any symmetry $G$ of the theory, which includes the case where $G$ is the entire effectively-acting symmetry.  This rules out $\io_G(\hat{G})$ as being an effectively-acting symmetry of $T$ -- in fact, it must be a trivially-acting symmetry.  However, there must be more to this story, as if we declared that $\io_G(\hat{G})$ acts trivially and left it at that, we would expect $T/(G\times\io_G(\hat{G}))$ to decompose into a sum of copies of $T/G$, whereas (\ref{qscomp}) tells us to instead expect a single, ungauged copy of $T$.

The trick is to remember that the trivial action of $\io_G(\hat{G})$ means that it comes with a set of TPOs, such that the full symmetry under consideration is
\be
(G_{[0]}\times\io_G(\hat{G})_{[0]}). \io_G(\hat{G})_{[1]}.
\ee
The missing ingredient is that there is in fact a mixed anomaly between $G_{[0]}$ and $\io_G(\hat{G})_{[1]}$, given by $\chi_{\hat{g}}(g)$.  It is in this way that $\hat{G}$, when `pulled back' to the symmetry $\io_G(\hat{G})$ of $T$, still `knows' about its action on twisted sector states.\footnote{In \cite[section 2.1]{QSDecomp} a similar setup was proposed where the phases entering the partition function arose from a choice of discrete torsion in the $G\times\io_G(\hat{G})$ orbifold.  Per the arguments in that paper, for a direct product of a trivially-acting and effectively acting symmetry, the phases arising from a mixed anomaly will always be equivalent to a choice of discrete torsion.  From the perspective advocated in this paper, it seems more correct to regard the source of such phases as a mixed anomaly.}  The mixed anomaly between 0-form and 1-form symmetries acts to obstruct decomposition in the 0-gauged theory, which is exactly what we see in (\ref{qscomp}) -- the naive decomposition that should arise from gauging a trivially-acting 0-form symmetry is not present.

Furthermore, we can turn this argument around.  The above established that $\io_G(\hat{G})$ is a symmetry of $T$; for notational clarity, let us write $\io_G(\hat{G})=I$.  This means that we could decide to ignore $G$ for the moment and consider $T/I$.  This gauged theory should have its own quantum symmetry, $\hat{I}$ such that $T/I/\hat{I}=T$.  We can now play the same game as before, by recognizing that there should be a symmetry $\io_I(\hat{I})$ of $T$ such that
\be
\label{qscomp2}
T/(I\times\io_I(\hat{I}))=T.
\ee
By comparing (\ref{qscomp}) and (\ref{qscomp2}), we can readily identify
\be
G=\io_{I}(\hat{I}).
\ee
In words, then, this exercise has taught us that to each (gaugable) effective symmetry $G$ in a theory $T$ there is a trivially-acting symmetry $I=\io_G(\hat{G})$ of $T$ which is the preimage of $G$'s quantum symmetry in $T/G$.  This trivially-acting preimage has its own dual quantum symmetry in $T/I$, and the preimage in $T$ of that symmetry is $G$ itself.

Let us say, for the sake of argument, that one wanted to avoid at all costs considering trivially-acting symmetries in one's field theories.  A takeaway of the above should be that trivially-acting symmetries are not just something one can introduce or not introduce at a whim -- they are intimately tied in with effectively-acting symmetries.  In particular, if one wants to take seriously the notion of orbifold composibility, then compositions involving quantum symmetries necessitate the introduction of trivially-acting symmetries.

\newpage

\section{Incorporating Lower-Form Symmetries}

We are now at the point where we have developed sufficient technology related to symmetries in $2d$ to discuss how $-1$-form symmetries enter the picture.  The exercises of section~\ref{sec:prelim} dealt with duals of 0-form symmetries and gaugings of 1-form symmetries, but we have so far said nothing about the duals to 1-form symmetries.  Of course this is not because the $-1$-form symmetries are inherently any more difficult to handle than 1-form symmetries -- rather, the presentation reflects the state of the literature.

We will begin with the simplest example: two copies of a theory $T$.  This system has two vacua (one for each copy), and therefore by selecting a projector we could gauge the 1-form symmetry.\footnote{As a reminder from the previous section, we could take a linear combination of the vacua that fuse as $\Z_2$ and call this a $\Z_2$ 1-form symmetry.  But at the end of the day, gauging this symmetry will simply correspond to selecting a projector, so imposing the $\Z_2$ group structure is entirely optional.}  This gauging produces a single copy of the theory $T$.  As already discussed extensively, this copy of $T$ should possess a quantum $-1$-form symmetry which we can gauge to recover the second copy.  This $-1$-form symmetry should be generated by two-dimensional topological operators, i.e.~spacetime-filling operators.

In the case at hand, let us write the initial, decomposing theory as
\be
T_+\oplus T_-
\ee
where the plus and minus markings are simply a way to label the two copies, without further connotation.  There are projection operators $\pi_+$ and $\pi_-$ which project onto $T_+$ and $T_-$, respectively, and these projectors sum to the identity.  This means that any correlation function in the $T_+\oplus T_-$ theory breaks up as
\be
\label{decompcfs}
\braket{\cdots} = \braket{1\cdots} = \braket{(\pi_++\pi_-)\cdots}=\braket{\pi_+\cdots}+\braket{\pi_-\cdots}.
\ee
We can gauge the 1-form symmetry by projecting onto, without loss of generality, $T_+$.  This is done by inserting $\pi_+$ into any correlation function, giving us
\be
\braket{\pi_+\cdots}=\braket{\pi_+\pi_+\cdots}+\braket{\pi_-\pi_+\cdots}=\braket{\pi_+\cdots},
\ee
which follows from the fusion relation of the projectors being $\pi_i\pi_j=\delta_{ij}\pi_i$.  Thus any correlation function in the gauged theory is truncated to its $T_+$ part.

Now we would like to undo this gauging.  In order to do this, we introduce a surface operator $\Pi$ which will serve as our `dual' projection operator (dual to the projector $\pi_+$, hence the similar notation).  It has the property that\footnote{Since $\pi_\pm$ only exist in the ungauged theory (i.e.~prior to gauging the 1-form symmetry), our definition of $\Pi$ and associated calculations will really be in terms of its preimage in the decomposing theory.  We will be sloppy in this section and use the same notation for $\Pi$ and its preimage.}
\be
\label{lambdaprop}
\pi_\pm\Pi=(\pi_++\pi_-)\id.
\ee
The meaning of this equation is that, in a region of our worldsheet filled with the surface operator $\Pi$, we can trade a $\pi_\pm$ operator living on this background for a ($\pi_++\pi_-)$ operator living on the trivial surface defect $\id$.  Let us see what effect such an operator would have on correlation functions.  Using $\Pi$ as the worldsheet background in our gauged theory, we have
\be
\braket{\pi_\pm\Pi\cdots}=\braket{(\pi_++\pi_-)\id\cdots}=\braket{\pi_+\cdots}+\braket{\pi_-\cdots}.
\ee
As promised, by inserting the surface operator $\Pi$, our correlation functions once again take the form (\ref{decompcfs}), and the 1-form gauging is undone!  That an operator with the properties of (\ref{lambdaprop}) should exist in a generic field theory seems surprising, but we will regard the existence of such an operator as equivalent to the assertion that the dual symmetry always exists.

While we would like to regard $\Pi$ as dual to the local projection operators, note that it is not quite a projector itself in the sense that 
\be
\pi_\pm\Pi^2=(\pi_++\pi_-)\Pi=2(\pi_++\pi_-)\id\neq\pi_\pm\Pi.
\ee
Intuitively, this is because we can only `project downwards' onto a single component of a sum once, but we can `project updwards' in the sense of creating additional copies of a theory as many times as we want.

Generalization of the above example is fairly immediate.  In $d$ spacetime dimensions, assume we have a direct sum of $N$ copies of a theory $T$.  We can choose a projector $\pi_j$ to project the theory onto one of the universes, gauging the associated $(d-1)$-form symmetry.  Gauging the dual $-1$-form symmetry will involve selecting a background $d$-dimensional operator $\Pi$ to fill the worldsheet.  $\Pi$ has the property that
\be
\pi_j\Pi=\left[\sum_{i=1}^N\pi_i\right]\id,
\ee
i.e.~the theory on the background $\Pi$ with an insertion of $\pi_j$ is equivalent to the theory on a trivial background with all projectors inserted.  Later in this section we will consider cases where the universes are not isomorphic.

\subsection{Reformulation as a Group-Like Gauging}

In the above treatment of gauging $-1$-form symmetries, we worked at the level of projectors and dual projectors/background fields.  However, per the discussions of section~\ref{sec:prelim}, we could have phrased the 1-form symmetry in the $2d$ example above as a $\Z_2$.  The expectation, then, is that its dual symmetry would also be a $\Z_2$.  Indeed, (\ref{h0t2}) implies that in such a formulation the $-1$-gauged partition function would be expressed as a sum over $\Z_2$.

Let us define a surface operator $\Sigma$ in the 1-gauged theory, as we did with $\Pi$, by its preimage in the ungauged theory:
\be
\label{sigmacond}
\pi_\pm\Sigma=\pi_\mp\id.
\ee
This $\Sigma$ is dual to the TPOs $\sigma$ of section~\ref{sec:prelim} in the same sense that $\Pi$ is dual to the projectors $\pi$.  $\Sigma$ naturally has a $\Z_2$ fusion algebra with $\id$, in the sense that
\begin{align}
\id\otimes\id&=\id, \\
\id\otimes\Sigma&=\Sigma,\\
\Sigma\otimes\id&=\Sigma,\\
\Sigma\otimes\Sigma&=\id.
\end{align}
We can check that this fusion is compatible with the definition of $\Sigma$ by calculating $\pi_\pm\Sigma\otimes\Sigma$ in two ways
\begin{align}
(\pi_\pm\Sigma)\otimes\Sigma&=\pi_\mp\id\otimes\Sigma=\pi_\mp\Sigma=\pi_\pm\id,\\
\pi_\pm(\Sigma\otimes\Sigma)&=\pi_\pm\id.
\end{align}
Thus, (the preimages of) $\id, \Sigma$ make $\pi_\pm$ into a $\Z_2$ torsor.  Now we can express $\Pi$ as
\be
\Pi=\id+\Sigma.
\ee
So, instead of obtaining correlation functions in the $-1$-gauged theory by inserting $\Pi$, we can obtain them as a sum
\be
\braket{\Pi\cdots}=\braket{\id\cdots}+\braket{\Sigma\cdots}
\ee
over a pair of operators with $\Z_2$ fusion, which recasts the $-1$-form gauging as a $\Z_2$ gauging.

Again, this picture immediately generalizes.  If we begin with $N$ copies of a theory in $d$ dimensions, there are $N$ vacua and $N$ projection operators $\pi_i$.  Label these vacua by irreducible representations $\hat{k}$ of a group $K$.\footnote{We are writing irreps of $K$ as $\hat{k}$, though these irreps only form a group $\hat{K}$ for $K$ abelian.  Even when $K$ is abelian, however, the labeling of the $\pi$ by elements of $\hat{K}$ is noncanonical; the $\pi_i$ have orthonormal fusion relations.}  In the 1-gauged theory, define a set of surface operators $\Sigma_{\hat{k}}$ labeled by elements of Rep($K$) by the requirement that the preimage of the $\Sigma_{\hat{k}}$ interact with the projectors as
\be
\pi_{\hat{k}_1}\Sigma_{\hat{k}_2}=\pi_{\hat{k}_1\otimes\hat{k}_2}\id.
\ee
For $K$ abelian, this gives the $\pi_i$ the structure of a $\hat{K}$ torsor (note that this definition implies $\Sigma_1=\id$).  Further, we have
\be
\Pi=\sum_{\hat{k}\in\text{Rep}(K)}\Sigma_{\hat{k}}.
\ee
Thus, gauging the $-1$-form symmetry can be written as a sum over irreps of $K$ via summing over insertions of $\Sigma_{\hat{k}}$ on the worldsheet.  This reproduces the expected structure of a $K_{[d-1]}$ symmetry with a $\text{Rep}(K)_{[-1]}$ dual.

As before, we stress that the $(d-1)$-form symmetry in question does not have any inherent group or fusion strcuture beyond the orthonormal relations of the projection operators (hence the non-canonical labeling of the $\pi_i$ by elements of Rep($K$)).  But we can choose to impose such a structure on it by regarding the $(d-1)$-form symmetry as arising from gauging a trivially-acting $(d-2)$-form symmetry.  Since the TPOs $\sigma_k$ are bound to the TDLs that generate the $(d-2)$-form symmetry in the ungauged theory, their fusion relations can be e.g.~non-abelian or non-group-like \cite{Sharpe_Noninvertible}.  The surface operators $\Sigma_{\hat{k}}$ defined here could then be regarded as the duals of these $\sigma_k$.

\subsection{Inhomogeneous Sums}

Now let us advance to the situation in which the initial direct sum is not over identical copies of a single theory.  For our first such example, consider a theory $T$ with a $\Z_2$ symmetry generated by $g$, and in turn consider the direct sum of $T$ with its $\Z_2$ orbifold:
\be
T\oplus T/\Z_2.
\ee
Let us define $\pi_+$ to be the projector associated to the unorbifolded copy $T$, similarly $\pi_-$ for $T/\Z_2$.  We will, as in previous examples, need a surface defect $\Sigma$.  In this case $\Sigma$ should act on the theory $T$ on a toroidal worldsheet as
\be
Z_{T}^\Sigma=\frac{1}{2}[Z_{1,1}+Z_{1,g}+Z_{g,1}+Z_{g,g}],
\ee
which is to say that we can trade an otherwise empty worldsheet in the $\Sigma$ background for a sum of worldsheets wrapped by TDLs for the $\Z_2$ symmetry.  This condition is pictured in Figure~\ref{fig:lfgauge}.  Similarly, $\Sigma$ should act on $T/\Z_2$ as
\be
Z_{T/\Z_2}^\Sigma=\frac{1}{2}[Z_{1,1}+Z_{1,\hat{g}}+Z_{\hat{g},1}+Z_{\hat{g},\hat{g}}],
\ee
where $\hat{g}$ generates the quantum $\Z_2$ symmetry that takes $T/\Z_2\to T$.

\begin{figure}
	\begin{tikzpicture}
	\draw[thin] (0.5,0)--(2.5,0);
	\draw[thin] (2.5,0)--(2.5,2);
	\draw[thin] (0.5,0)--(0.5,2);
	\draw[thin] (0.5,2)--(2.5,2);
	\fill[gray!20,opacity=0.5] (0.5,0) -- (0.5,2) -- (2.5,2) -- (2.5,0) -- cycle;
	\node at (1.5,1) {$\Sigma$};
	\node at (3,1) {$=$};
	\node at (3.75,1) {$\frac{1}{2}$};
	\draw[thin] (4,0)--(6,0);
	\draw[thin] (6,0)--(6,2);
	\draw[thin] (4,0)--(4,2);
	\draw[thin] (4,2)--(6,2);
	\node at (6.5,1) {$+\text{ }\frac{1}{2}$};
	\draw[thin] (7,0)--(9,0);
	\draw[thin] (9,0)--(9,2);
	\draw[thin] (7,0)--(7,2);
	\draw[thin] (7,2)--(9,2);
	\draw[thick,->] (8,0) -- (8,1);
	\draw[thick] (8,1) -- (8,2);
	\node at (8.4,1.4) {$g$};
	\node at (9.5,1) {$+\text{ }\frac{1}{2}$};
	\draw[thin] (10,0)--(12,0);
	\draw[thin] (12,0)--(12,2);
	\draw[thin] (10,0)--(10,2);
	\draw[thin] (10,2)--(12,2);
	\draw[thick,->] (10,1)--(11,1);
	\draw[thick] (11,1)--(12,1);
	\node at (10.6,1.4) {$g$};
	\node at (12.5,1) {$+\text{ }\frac{1}{2}$};
	\draw[thin] (13,0)--(15,0);
	\draw[thin] (15,0)--(15,2);
	\draw[thin] (13,0)--(13,2);
	\draw[thin] (13,2)--(15,2);
	\draw[thick,->] (14,0)--(14,1.5);
	\draw[thick] (14,1.5)--(14,2);
	\draw[thick,->] (13,1)--(13.5,1);
	\draw[thick] (13.5,1)--(15,1);
	\node at (14.4,1.6) {$g$};
	\node at (13.4,0.6) {$g$};
	\end{tikzpicture}
	\caption{The $\Sigma$ background for $T$.}
	\label{fig:lfgauge}
\end{figure}

Clearly such a $\Sigma$ has the effect of swapping the universes, so we could equally well phrase its action as (\ref{sigmacond}), matching our earlier example.  Then once again we have the dual projector
\be
\Pi=1+\Sigma,
\ee
and inserting $\Pi$ gauges the $-1$-form symmetry.\\

In our next example, the theories in question differ by discrete torsion.  We begin with
\be
\label{dtex}
\left[\frac{T}{\Z_2\times\Z_2}\right]_{1}\oplus\left[\frac{T}{\Z_2\times\Z_2}\right]_{\omega}
\ee
where on the left we have taken the trivial cocycle in $H^2(\Z_2\times\Z_2,U(1))$ as our choice of discrete torsion, while on the right we take a representative $\omega$ of the non-trivial class.  Similarly to the previous examples, we can define $\pi_+$ to project onto the theory on the left and $\pi_-$ onto the right.

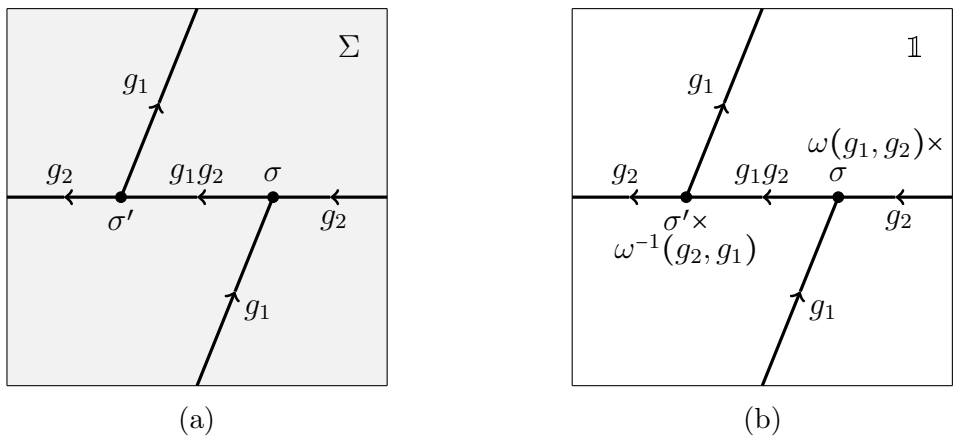
\begin{figure}
	\begin{subfigure}{0.5\textwidth}
	\centering
	\begin{tikzpicture}
	\fill[gray!20,opacity=0.5] (0,0) -- (0,5) -- (5,5) -- (5,0) -- cycle;
	\draw[thin] (0,0)--(5,0);
	\draw[thin] (5,0)--(5,5);
	\draw[thin] (0,0)--(0,5);
	\draw[thin] (0,5)--(5,5);
	\draw[very thick,->] (2.5,0)--(3,1.25);
	\draw[very thick] (3,1.25)--(3.5,2.5);
	\draw[very thick,->] (1.5,2.5)--(2,3.75);
	\draw[very thick] (2,3.75)--(2.5,5);
	\draw[very thick,->] (5,2.5)--(4.25,2.5);
	\draw[very thick] (4.25,2.5)--(3.5,2.5);
	\draw[very thick,->] (3.5,2.5)--(2.5,2.5);
	\draw[very thick] (2.5,2.5)--(1.5,2.5);
	\draw[very thick,->] (1.5,2.5)--(0.75,2.5);
	\draw[very thick] (0.75,2.5)--(0,2.5);
	\node at (1.7,4) {$g_1$};
	\node at (3.3,1) {$g_1$};
	\node at (0.7,2.8) {$g_2$};
	\node at (4.3,2.2) {$g_2$};
	\node at (2.5,2.8) {$g_1g_2$};
	\node at (4.5,4.5) {$\Sigma$};
	\filldraw[black] (1.5,2.5) circle (2pt);
	\node at (1.5,2.2) {$\sigma'$};
	\filldraw[black] (3.5,2.5) circle (2pt);
	\node at (3.5,2.8) {$\sigma$};
	\end{tikzpicture}
	\caption{}
	\label{lfdtex1}
	\end{subfigure}
	\begin{subfigure}{0.5\textwidth}
	\centering
	\begin{tikzpicture}
	\draw[thin] (0,0)--(5,0);
	\draw[thin] (5,0)--(5,5);
	\draw[thin] (0,0)--(0,5);
	\draw[thin] (0,5)--(5,5);
	\draw[very thick,->] (2.5,0)--(3,1.25);
	\draw[very thick] (3,1.25)--(3.5,2.5);
	\draw[very thick,->] (1.5,2.5)--(2,3.75);
	\draw[very thick] (2,3.75)--(2.5,5);
	\draw[very thick,->] (5,2.5)--(4.25,2.5);
	\draw[very thick] (4.25,2.5)--(3.5,2.5);
	\draw[very thick,->] (3.5,2.5)--(2.5,2.5);
	\draw[very thick] (2.5,2.5)--(1.5,2.5);
	\draw[very thick,->] (1.5,2.5)--(0.75,2.5);
	\draw[very thick] (0.75,2.5)--(0,2.5);
	\node at (1.7,4) {$g_1$};
	\node at (3.3,1) {$g_1$};
	\node at (0.7,2.8) {$g_2$};
	\node at (4.3,2.2) {$g_2$};
	\node at (2.5,2.8) {$g_1g_2$};
	\node at (4.5,4.5) {$\id$};
	\filldraw[black] (1.5,2.5) circle (2pt);
	\node at (1.5,2.2) {$\sigma'\times$};
	\node at (1.5,1.8) {$\omega^{-1}(g_2,g_1)$};
	\filldraw[black] (3.5,2.5) circle (2pt);
	\node at (3.5,2.8) {$\sigma$};
	\node at (4,3.2) {$\omega(g_1,g_2)\times$};
	\end{tikzpicture}
	\caption{}
	\label{lfdtex2}
	\end{subfigure}
	\caption{The effect of $\Sigma$ on a $\Z_2\times\Z_2$ gauge bundle over a toroidal worldsheet for the theory $T$.}
	\label{fig:lfdtex}
\end{figure}

As before we would like to define a background $\Sigma$ which swaps the theories.  In this case $\Sigma$ is most conveniently defined not by its preimage in the decomposing theory, but further by its preimage under the $\Z_2\times\Z_2$ gauging.  Following the treatment of discrete torsion in section~\ref{sec:gaugeone}, the cost of swapping from the $\Sigma$ background to the trivial background $\id$ will be to multiply the operator sitting at a three-way junction where $g_1$ and $g_2$ meet $g_1g_2$ by $\omega(g_1,g_2)$.  This procedure is pictured in Figure~\ref{fig:lfdtex}, where Figure~\ref{lfdtex1} shows a network of TDLs over a toroidal worldsheet in a $\Sigma$ background.  The operators at the two three-way junctions are labeled by $\sigma$ and $\sigma'$.  Swapping $\Sigma$ for the trivial background $\id$ leads to Figure~\ref{lfdtex2}, in which $\sigma$ has received the factor $\omega(g_1,g_2)$ and $\sigma'$ has received $\omega^{-1}(g_2,g_1)$.  When we sum over such TDL insertions to build the orbifold partition function, this process changes the summand by by $\omega(g_1,g_2)/\omega(g_2,g_1)$, making it clear that the $\Sigma$ background effected a swap of discrete torsion.

Insertion of $\Pi=1+\Sigma$ recovers the direct sum (\ref{dtex}) beginning from either of the universes.  Note the strong similarity of this example to our initial quantum mechanical example of section~\ref{sec:qm}.  There we had a decomposition into two theories that differed by a discrete theta angle, and we projected onto a single universe.  In order to recover the initial direct sum theory, we summed over a trivial background TDL (the analogue of $\id$) and a non-trivial one which implemented a theta angle shift, effectively swapping the universes (as does $\Sigma$).  This comparison helps to demonstrate that the manipulation of dual pairs of $(d-1)$-form and $-1$-form symmetries is fairly generic throughout dimensions.

\subsection{A More Complete Picture}

With basic results regarding $-1$-form symmetries worked out, we can give a schematic treatment of the symmetries enjoyed by direct sum theories.  Let us begin with a theory $T$, which we will regard as having a trivially-acting global symmetry $K$.  For simplicity we will take $K$ to be abelian, though it is not necessary.  As we have discussed at length, $K$ should be regarded as a mix of a 0-form and 1-form symmetry, which in turn is controlled by a mix of topological line and point operators.  We would write the symmetry of $T$ as $K_{[0]}. K_{[1]}$, where the extension class is the identity morphism $\delta$ of $K$.

We can gauge $K_{[0]}$ to arrive at a decomposing theory with symmetry $\hat{K}_{[0]}\times K_{[1]}$.  This gauged theory has $|K|$ vacua which can be constructed as linear combinations of the TPOs from the ungauged theory.  The new 0-form symmetry $\hat{K}_{[0]}$ is an effective symmetry which acts on an operator in the $k$-twisted sector as $\chi_{\hat{k}}(k)$.  This immediately tells us that there is a mixed anomaly between the symmetries $\hat{K}_{[0]}$ and $K_{[1]}$, which shows up as the phase $\chi_{\hat{k}}(k)$ in the action of a TDL labeled by $\hat{k}$ on a TPO $\sigma_k$.  The decomposing theory $T/K_{[0]}$ can thus be gauged by $\hat{K}_{[0]}$, which would return us to $T$, or by $K_{[1]}$ -- the mixed anomaly obstructs gauging these symmetries simultaneously.

We could then gauge the 1-form symmetry of $T/K_{[0]}$ to once again obtain a single copy of $T$, except this time we regard it as having a quantum $-1$-form symmetry arising from the 1-form gauging.  The mixed anomaly of $T/K_{[0]}$ will once again become an extension class, this time preventing $\hat{K}_{[0]}$ from constituting a standalone symmetry.  This is the theory we should have obtained by gauging $T$ by its full trivially-acting symmetry $K_{[0]}. K_{[1]}$.  The results of this symmetry yoga are tabulated in Figure~\ref{fig:symmtable}.

\begin{figure}[h]
	\begin{tabular}{l l l l l}
		Theory & Symmetry & Ext.~Class & Mixed Anomaly & Allowed Gaugings\\
		\hline
		$T$ & $K_{[0]}. K_{[1]}$ & $\delta:K\to K$ & Trivial & $K_{[0]}$, $K_{[0]}. K_{[1]}$\\
		$T/K_{[0]}$ & $\hat{K}_{[0]} \times K_{[1]}$ & Trivial & $\chi_{\hat{k}}(k)$ & $\hat{K}_{[0]}$, $K_{[1]}$\\
		$T/(K_{[0]}. K_{[1]})$ & $\hat{K}_{[-1]}.\hat{K}_{[0]}$ & $\tilde{\delta}:\hat{K}\to\hat{K}$ & Trivial & $\hat{K}_{[-1]}$, $\hat{K}_{[-1]}.\hat{K}_{[0]}$
	\end{tabular}
	\caption{Various gaugings of a single 2d theory without taking any effective symmetries into account.}
	\label{fig:symmtable}
\end{figure}

In particular we see that each of the three theories listed here admit two gaugings (not including possible subgroups), which tells us that we can move from any one of these theories to any other one.  It is interesting to ponder whether $T$ and $T/(K_{[0]}. K_{[1]})$ should be considered the same theory.  Certainly they have the same spectrum of local operators, and furthremore have the same partition function on a Riemann surface of any genus.  Their description in terms of extended operators, i.e.~the symmetries they carry, is what differs (of course for an effective symmetry this sentence and the previous would be at odds, as that symmetry would need to manifest itself somewhere in the correlation functions of the theory).  Specifically, we are regarding $T$ as carrying a 0-form symmetry (which acts trivially), while $T/(K_{[0]}. K_{[1]})$ does not have any such symmetry.  It would seem reasonable, at the very least, to regard $T$ and $T/(K_{[0]}. K_{[1]})$ as dual descriptions of the same theory.\\

Now consider that $T$ may have an effectively-acting 0-form symmetry $G$, such that its total 0-form symmetry is $\Gamma_{[0]}=K_{[0]}. G_{[0]}$.  Again we will use notation suggestive of $\Gamma$ being abelian, though similar statements hold with suitable generalization in the non-abelian case.  The extension class is a homomorphism from $K$ to $\Gamma$, which is given by inclusion as a subgroup.  When gauging $\Gamma_{[0]}$, the mixed anomaly now needs to prescribe a phase to the action of a line labeled by $\hat{\Gamma}$ on a local operator labeled by $k$.  This is done with a character of $\Gamma$ in combination with the extension class, which leads to a mixed anomaly we can write as $\chi_{\hat{\gamma}}(\delta(k))$.

Gauging the 1-form symmetry $K_{[1]}$ of the resulting theory brings us back to a single copy, but this time we would say its symmetry is $\hat{K}_{[-1]}.\hat{\Gamma}_{[0]}$.  Here the non-trivial extension class serves to bind the lines corresponding to the trivial part of $\hat{\Gamma}_{[0]}$ to the surfaces which generate $\hat{K}_{[-1]}$.  Following (\ref{mixedextclass}), this extension class should be valued in $H^1(\mathcal{K}(\hat{\Gamma},1),\hat{K})=H^1(B\hat{\Gamma},\hat{K})=\text{Hom}(\hat{\Gamma},\hat{K})$.  If we were to gauge $\hat{K}_{[-1]}$ to return to the decomposing theory, we would make a different prediction for its mixed anomaly -- it should be given by a character $\tilde{\chi}$ of $\hat{K}$, and the phase appearing in the action of a line labeled by $\hat{\gamma}$ on a local operator labeled by $k$ would be $\tilde{\chi}_{k}(\tilde{\delta}(\hat{\gamma}))$.  These two descriptions of the mixed anomaly should agree, so we should have
\be
\chi_{\hat{\gamma}}(\delta(k))=\tilde{\chi}_{k}(\tilde{\delta}(\hat{\gamma})),
\ee
which can be seen as a consistency condition on the two extension classes $\delta$ and $\tilde{\delta}$.  Figure~\ref{fig:symmtable2} summarizes the results of this exercise.

\begin{figure}[h]
	\begin{tabular}{l l l l l}
		Theory & Symmetry & Ext.~Class & Mixed Anomaly & Allowed Gaugings\\
		\hline
		$T$ & $\Gamma_{[0]}. K_{[1]}$ & $\delta:K\to \Gamma$ & Trivial & $\Gamma_{[0]}$, $\Gamma_{[0]}. K_{[1]}$\\
		$T/\Gamma_{[0]}$ & $\hat{\Gamma}_{[0]} \times K_{[1]}$ & Trivial & $\chi_{\hat{\gamma}}(\delta(k))=\tilde{\chi}_{k}(\tilde{\delta}(\hat{\gamma}))$ & $\hat{\Gamma}_{[0]}$, $K_{[1]}$\\
		$T/(\Gamma_{[0]}. K_{[1]})$ & $\hat{K}_{[-1]}.\hat{\Gamma}_{[0]}$ & $\tilde{\delta}:\hat{\Gamma}\to\hat{K}$ & Trivial & $\hat{K}_{[-1]}$, $\hat{K}_{[-1]}.\hat{\Gamma}_{[0]}$
	\end{tabular}
	\caption{Various gaugings of a single 2d theory with effective symmetry $G_{[0]}$ and trivially-acting symmetry $K_{[0]}$ combining as $\Gamma_{[0]}=K_{[0]}. G_{[0]}$.}
	\label{fig:symmtable2}
\end{figure}


\section{Conclusion}

We now have a clearer picture of the relationship between 0-dimensional and $d$-dimensional topological operators, which are dual under gauging.  Making this relationship explicit has involved the consideration of non-trivial background operators which fill the worldsheet of our theory.  In return, we are able to settle on a more complete view of symmetries in field theories, in which we can verify the existence of a quantum dual to any given symmetry.  We can also more readily extend such considerations to mixes between symmetries controlled by topological operators of differing dimension.

There are certainly open questions that follow from these considerations.  One might hope for a clearer picture of how gauging a $-1$-form symmetry relates to the operation of coupling to a scalar background gauge field.  Also, the story told here should relate to the $-1$-branes of string theory \cite{Witten}, and such connections would be interesting to pursue.  Additionally, the spacetime-filling background operators that implement $-1$-form symmetries could be studied in their own right.  Appendix~\ref{sec:spt} presents a bare-bones interpretation of these background fields in terms of defects in a higher dimensional bulk theory, but further work is required to flesh out such a picture.  These few questions represent a small selection of the many mysteries that litter the path to a more complete understanding of symmetry in field theory.

\section*{Acknowledgments}

The author would like to thank Daniel Robbins and Eric Sharpe for helpful comments on a draft of this paper.

\appendix

\section{Bulk Interpretation}
\label{sec:spt}

In this appendix we will briefly sketch how the spcetime-filling operators that implement a $-1$-form symmetry can be understood when our $d$-dimensional system is viewed as living on the boundary of a $(d+1)$-dimensional system.

\begin{figure}
	\begin{subfigure}{0.5\textwidth}
	\centering
	\begin{tikzpicture}
	\draw[thick,dashed] (0,0) [partial ellipse=0:360:2cm and 0.5cm];
	\draw[thick,dashed] (-2,0) -- (-2,-2);
	\draw[thick,dashed] (2,0) -- (2,-2);
	\draw[thick] plot [smooth,tension=1.8] coordinates {(0.25,-1.25)  (0,-0.7) (0.05,-0.5)};
	\draw[thick,->] (1,-2) -- (0.5, -1.5);
	\draw[thick] (0.5,-1.5) -- (0.25,-1.25);
	\filldraw[black] (-0.025,-0.5) circle (2pt);
	\node at (0,-0.25) {$\sigma_g$};
	\node at (0.6,-1.1) {$\mathcal{L}_g$};
	\end{tikzpicture}
	\caption{}
	\label{spt1}
	\end{subfigure}
	\begin{subfigure}{0.5\textwidth}
	\centering
	\begin{tikzpicture}
	\draw[thick,->] (0,0) [partial ellipse=-90:90:2cm and 0.5cm];
	\draw[thick,->] (0,0) [partial ellipse=90:270:2cm and 0.5cm];
	\draw[thick,dashed] (-2,0) -- (-2,-2);
	\draw[thick,dashed] (2,0) -- (2,-2);
	\node at (0,-0.25) {$\mathcal{L}_{\hat{g}}$};
	\end{tikzpicture}
	\caption{}
	\label{spt2}
	\end{subfigure}
\caption{A coupled 1/2d system, before and after gauging.}
\label{fig:spt}
\end{figure}
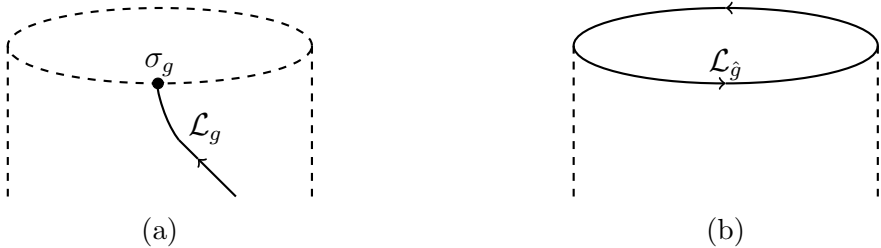

Let us begin with a 2d bulk having a 1d boundary.  For concreteness, we will take the 2d theory to live on a cylinder, with the 1d theory on one of its bounding circles -- this discussion will then map nicely to the example studied in section~\ref{sec:qm}.  As discussed in that section, a symmetry $G$ of the boundary theory is implemented by the insertion of local operators $\sigma_g$ labeled by elements of $G$.  We take the bulk theory to also have $G$ symmetry, implemented by TDLs $\mathcal{L}_g$.  Each insertion of $\sigma_g$ on the boundary connnects to a line $\mathcal{L}_g$ in the bulk, shown in Figure~\ref{spt1}.  We assign Neumann boundary conditions to the lines, such that the $\sigma_g$ can be freely moved around the boundary, but the lines must always end on a single boundary point.  

The boundary $G$ symmetry could carry a gauge anomaly corresponding to a class $\varepsilon\in H^2(G,U(1))$, which would essentially correspond to discrete torsion in the bulk TDL network, in the sense of section~\ref{sec:gaugeone}.  Since we plan on gauging the symmetry in question, we will assume that $\varepsilon$ is trivial in cohomology.  The partition function of the 2d theory with a given TDL network is given by the product of the phases $\varepsilon$ assigned to each junction (once again see section~\ref{sec:gaugeone}, in particular Figure~\ref{fig:dt_swap}).  Such a bulk theory is known as a Symmetry Protected Topological (SPT) phase \cite{Wen,Tachikawa_Lectures}.

Now we would like to gauge $G$ on both the bulk and the boundary.  Gauging the SPT in the bulk produces a TQFT, namely Dijkgraaf-Witten theory \cite{DW}.  The relevant fact for us is that the resulting theory has a  quantum symmetry, and there are now bulk TDLs $\mathcal{L}_{\hat{g}}$ for $\hat{g}\in\text{Rep}(G)$ corresponding to it.  In contrast to the TDLs for $G$, this new set will not have boundary conditions constraining them to a perpendicular intersection with the boundary -- in fact, these new lines can lie parallel to the boundary, and thus are localizable to it, the result of which is illustrated in Figure~\ref{spt2}.  Thus, we interpret the 1-dimensional operators that implement the boundary $-1$-form symmetry as the bulk TDLs $\mathcal{L}_{\hat{g}}$ pushed onto the boundary (and certainly gauging the bulk Rep$(G)$ must involve all such configurations).

This picture readily generalizes.  Assume we have a $d$-dimensional theory with a $(d-1)$-form symmetry corresponding an abelian group $G$.  Such a symmetry is implemented by local operators.  Regard this theory as living on the boundary of a $(d+1)$-dimensional theory.  The bulk theory should then have a corresponding $(d-1)$-form symmetry -- but this is now a $(d-1)$-form symmetry in $d+1$ dimensions, which means it is controlled by operators of codimension $d$, i.e.~TDLs.  We then have the same story as above where local operators on the boundary connect to TDLs with Neumann boundary conditions in the bulk.

Now we gauge the bulk $(d-1)$-form symmetry.  The quantum dual to a $(d-1)$-form symmetry in $d+1$ dimensions is a 0-form symmetry.  The resulting theory then has operators of dimension $d$ which implement the 0-form symmetry $\hat{G}$.  These operators can once again be localized to the boundary, where they will fill the entire $d$-dimensional worldsheet, implementing the dual $-1$-form symmetry to the boundary theory's original $(d-1)$-form symmetry.  In this way the quantum symmetries of the bulk and boundary systems are in complete correspondence.


\addcontentsline{toc}{section}{References}

\bibliographystyle{utphys}
\bibliography{LowerFormSymmetries}

\providecommand{\href}[2]{#2}\begingroup\raggedright\begin{thebibliography}{10}

\bibitem{Vafa}
C.~Vafa, ``{Quantum Symmetries of String Vacua},''
  \href{http://dx.doi.org/10.1142/S0217732389001842}{{\em Mod. Phys. Lett. A}
  {\bfseries 4} (1989) 1615}.

\bibitem{GaiottoKapustinSeibergWillett}
D.~Gaiotto, A.~Kapustin, N.~Seiberg, and B.~Willett, ``Generalized global
  symmetries,'' \href{http://dx.doi.org/10.1007/jhep02(2015)172}{{\em Journal
  of High Energy Physics} {\bfseries 2015} no.~2, (Feb, 2015) }.
  \url{https://doi.org/10.1007%2Fjhep02%282015%29172}.

\bibitem{Yu}
M.~Yu, ``Symmetries and anomalies of (1+1)d theories: 2-groups and symmetry
  fractionalization,'' \href{http://dx.doi.org/10.1007/jhep08(2021)061}{{\em
  Journal of High Energy Physics} {\bfseries 2021} no.~8, (Aug, 2021) }.
  \url{https://doi.org/10.1007%2Fjhep08%282021%29061}.

\bibitem{CordovaFreedLamSeiberg}
C.~Cordova, D.~Freed, H.~T. Lam, and N.~Seiberg, ``Anomalies in the space of
  coupling constants and their dynamical applications i,''
  \href{http://dx.doi.org/10.21468/scipostphys.8.1.001}{{\em {SciPost} Physics}
  {\bfseries 8} no.~1, (Jan, 2020) }.
  \url{https://doi.org/10.21468%2Fscipostphys.8.1.001}.

\bibitem{KapustinSeiberg}
A.~Kapustin and N.~Seiberg, ``Coupling a {QFT} to a {TQFT} and duality,''
  \href{http://dx.doi.org/10.1007/jhep04(2014)001}{{\em Journal of High Energy
  Physics} {\bfseries 2014} no.~4, (Apr, 2014) }.
  \url{https://doi.org/10.1007%2Fjhep04%282014%29001}.

\bibitem{FuchsRunkelSchweigert}
J.~Fuchs, I.~Runkel, and C.~Schweigert, ``{TFT} construction of {RCFT}
  correlators i: partition functions,''
  \href{http://dx.doi.org/10.1016/s0550-3213(02)00744-7}{{\em Nuclear Physics
  B} {\bfseries 646} no.~3, (Dec, 2002) 353--497}.
  \url{https://doi.org/10.1016%2Fs0550-3213%2802%2900744-7}.

\bibitem{AndoHellermanHenriquesPantevSharpe}
M.~Ando, S.~Hellerman, A.~Henriques, T.~Pantev, and E.~Sharpe, ``Cluster
  decomposition, t-duality, and gerby {CFTs},''
  \href{http://dx.doi.org/10.4310/atmp.2007.v11.n5.a2}{{\em Advances in
  Theoretical and Mathematical Physics} {\bfseries 11} no.~5, (2007) 751--818}.
  \url{https://doi.org/10.4310%2Fatmp.2007.v11.n5.a2}.

\bibitem{PantevRobbinsSharpeVandermeulen}
T.~Pantev, D.~G. Robbins, E.~Sharpe, and T.~Vandermeulen, ``Orbifolds by
  2-groups and decomposition,''
  \href{http://dx.doi.org/10.1007/jhep09(2022)036}{{\em Journal of High Energy
  Physics} {\bfseries 2022} no.~9, (Sep, 2022) }.
  \url{https://doi.org/10.1007%2Fjhep09%282022%29036}.

\bibitem{TopOps}
D.~G. Robbins, E.~Sharpe, and T.~Vandermeulen, ``Decomposition,
  trivially-acting symmetries, and topological operators,'' 2022.
\newblock In preparation.

\bibitem{BhardwajTachikawa}
L.~Bhardwaj and Y.~Tachikawa, ``On finite symmetries and their gauging in two
  dimensions,'' \href{http://dx.doi.org/10.1007/jhep03(2018)189}{{\em Journal
  of High Energy Physics} {\bfseries 2018} no.~3, (Mar, 2018) }.
  \url{https://doi.org/10.1007%2Fjhep03%282018%29189}.

\bibitem{BrunnerCarquevillePlencner}
I.~Brunner, N.~Carqueville, and D.~Plencner, ``Discrete torsion defects,''
  \href{http://dx.doi.org/10.1007/s00220-015-2297-9}{{\em Communications in
  Mathematical Physics} {\bfseries 337} no.~1, (Feb, 2015) 429--453}.
  \url{https://doi.org/10.1007%2Fs00220-015-2297-9}.

\bibitem{FrohlichFuchsRunkelSchweigert}
J.~Frohlich, J.~Fuchs, I.~Runkel, and C.~Schweigert,
  \href{http://dx.doi.org/10.1142/9789814304634_0056}{``{Defect lines,
  dualities, and generalised orbifolds},''} in {\em {16th International
  Congress on Mathematical Physics}}.
\newblock 9, 2009.
\newblock \href{http://arxiv.org/abs/0909.5013}{{\ttfamily arXiv:0909.5013
  [math-ph]}}.

\bibitem{Wen}
X.-G. Wen, ``Classifying gauge anomalies through symmetry-protected trivial
  orders and classifying gravitational anomalies through topological orders,''
  \href{http://dx.doi.org/10.1103/physrevd.88.045013}{{\em Physical Review D}
  {\bfseries 88} no.~4, (Aug, 2013) }.
  \url{https://doi.org/10.1103%2Fphysrevd.88.045013}.

\bibitem{GaiottoKulp}
D.~Gaiotto and J.~Kulp, ``Orbifold groupoids,''
  \href{http://dx.doi.org/10.1007/jhep02(2021)132}{{\em Journal of High Energy
  Physics} {\bfseries 2021} no.~2, (Feb, 2021) }.
  \url{https://doi.org/10.1007%2Fjhep02%282021%29132}.

\bibitem{GaiottoKapustinKomargodskiSeiberg}
D.~Gaiotto, A.~Kapustin, Z.~Komargodski, and N.~Seiberg, ``Theta, time reversal
  and temperature,'' \href{http://dx.doi.org/10.1007/jhep05(2017)091}{{\em
  Journal of High Energy Physics} {\bfseries 2017} no.~5, (May, 2017) }.
  \url{https://doi.org/10.1007%2Fjhep05%282017%29091}.

\bibitem{KikuchiTanizaki}
Y.~Kikuchi and Y.~Tanizaki, ``Global inconsistency, 't hooft anomaly, and level
  crossing in quantum mechanics,''
  \href{http://dx.doi.org/10.1093/ptep/ptx148}{{\em Progress of Theoretical and
  Experimental Physics} {\bfseries 2017} no.~11, (Nov, 2017) }.
  \url{https://doi.org/10.1093%2Fptep%2Fptx148}.

\bibitem{Sharpe_Diverse}
E.~Sharpe, ``Decomposition in diverse dimensions,''
  \href{http://dx.doi.org/10.1103/physrevd.90.025030}{{\em Physical Review D}
  {\bfseries 90} no.~2, (Jul, 2014) }.
  \url{https://doi.org/10.1103%2Fphysrevd.90.025030}.

\bibitem{Tachikawa}
Y.~Tachikawa, ``On gauging finite subgroups,''
  \href{http://dx.doi.org/10.21468/scipostphys.8.1.015}{{\em {SciPost} Physics}
  {\bfseries 8} no.~1, (Jan, 2020) }.
  \url{https://doi.org/10.21468%2Fscipostphys.8.1.015}.

\bibitem{ATLAS}
R.~A. Wilson, J.~H. Conway, and S.~P. Norton, ``Atlas of finite groups,''
\newblock 1985.

\bibitem{ChangLinShaoYin}
C.-M. Chang, Y.-H. Lin, S.-H. Shao, Y.~Wang, and X.~Yin, ``Topological defect
  lines and renormalization group flows in two dimensions,''
  \href{http://dx.doi.org/10.1007/jhep01(2019)026}{{\em Journal of High Energy
  Physics} {\bfseries 2019} no.~1, (Jan, 2019) }.
  \url{https://doi.org/10.1007%2Fjhep01%282019%29026}.

\bibitem{Ginsparg}
P.~Ginsparg, ``Applied conformal field theory,''.
  \url{https://arxiv.org/abs/hep-th/9108028}.

\bibitem{Tachikawa_Lectures}
Y.~Tachikawa, ``Topological phases and relativistic qfts,'' 2018.
\newblock
  \url{https://member.ipmu.jp/yuji.tachikawa/lectures/2018-cern-rikkyo/}.

\bibitem{Sharpe_UndoingDecomposition}
E.~Sharpe, ``Undoing decomposition,''
  \href{http://dx.doi.org/10.1142/s0217751x19502336}{{\em International Journal
  of Modern Physics A} {\bfseries 34} no.~35, (Dec, 2019) 1950233}.
  \url{https://doi.org/10.1142%2Fs0217751x19502336}.

\bibitem{Sharpe_Noninvertible}
E.~Sharpe, ``Topological operators, noninvertible symmetries and
  decomposition,'' 2021.
\newblock \url{https://arxiv.org/abs/2108.13423}.

\bibitem{QSDecomp}
D.~G. Robbins, E.~Sharpe, and T.~Vandermeulen, ``Quantum symmetries in
  orbifolds and decomposition,''
  \href{http://dx.doi.org/10.1007/jhep02(2022)108}{{\em Journal of High Energy
  Physics} {\bfseries 2022} no.~2, (Feb, 2022) }.
  \url{https://doi.org/10.1007%2Fjhep02%282022%29108}.

\bibitem{Witten}
E.~Witten, ``D-branes and k-theory,''
  \href{http://dx.doi.org/10.1088/1126-6708/1998/12/019}{{\em Journal of High
  Energy Physics} {\bfseries 1998} no.~12, (Dec, 1998) 019--019}.
  \url{https://doi.org/10.1088%2F1126-6708%2F1998%2F12%2F019}.

\bibitem{DW}
R.~Dijkgraaf and E.~Witten, ``{Topological gauge theories and group
  cohomology},'' \href{http://dx.doi.org/cmp/1104180750}{{\em Communications in
  Mathematical Physics} {\bfseries 129} no.~2, (1990) 393 -- 429}.
  \url{https://doi.org/}.

\end{thebibliography}\endgroup

\end{document}